\documentclass[fleqn,usenatbib]{mnras}
\usepackage{graphicx}
\usepackage[usenames,dvipsnames]{color}
\usepackage{times,ulem}
\usepackage{bm}
\usepackage{amsmath,amssymb}
\usepackage{fixltx2e} 

\newcommand{\dd}{\mathrm{d}}    		
\newcommand{\Brms}{\mathcal{B}_0}    		
\newcommand{\Db}{D_b}    		

\hypersetup{
pdftitle={Global diffusion of cosmic rays in random magnetic fields},
pdfauthor={
            A.~P.~Snodin (orcid.org/?),
            A.~Shukurov (orcid.org/0000-0001-6200-4304),
            G.~R.~Sarson (orcid.org/?),
            P.~J.~Bushby (orcid.org/?),
            L.~F.~S.~Rodrigues (orcid.org/0000-0002-3860-0525)
          }
}

\title[Global diffusion of cosmic rays]
{Global diffusion of cosmic rays in random magnetic fields}
\author[A.~P.~Snodin, A.~Shukurov, G.~R.~Sarson, P.~J.~Bushby and L.~F.~S.~Rodrigues]
{\parbox{\textwidth}{A.~P.~Snodin,$^{1,2}$
\thanks{E-mail: andrew.snodin@gmail.com; anvar.shukurov@ncl.ac.uk; \newline g.r.sarson@ncl.ac.uk; paul.bushby@ncl.ac.uk; luiz.\-ro\-dri\-gues@ncl.ac.uk}
A.~Shukurov,$^3$ G.~R.~Sarson,$^3$ P.~J.~Bushby$^3$ and L.~F.~S.~Rodrigues$^3$}\vspace{0.4cm}\\
\parbox{\textwidth}{$^1$Department of Materials and Production Technology Engineering,
Faculty of Engineering, King Mongkut's University of Technology North Bangkok, \\
Bangkok 10800, Thailand\\
$^2$Department of Mathematics, Faculty of Applied Science, King Mongkut's University
of Technology North Bangkok, Bangkok 10800, Thailand\\
$^3$School of Mathematics and Statistics, Newcastle University,
Newcastle upon Tyne, NE1 7RU, U.K.}}

\date{\today}

\renewcommand{\vec}[1]{\bmath{#1}}

\newcommand{\EQ}{\begin{equation}}
\newcommand{\EN}{\end{equation}}
\newcommand{\EQA}{\begin{eqnarray}}
\newcommand{\ENA}{\end{eqnarray}}

\newcommand{\Bhat}{\widehat{B}}
\newcommand{\BBhat}{\widehat{\vec{B}}}
\newcommand{\BB}{{\vec{B}}}
\newcommand{\larmor}{{_\mathrm{L}}}
\newcommand{\RL}{R_\mathrm{L}}
\newcommand{\lc}{{l_{\mathrm{c}}}}
\newcommand{\kmax}{{k_\mathrm{max}}}

\newcommand{\kres}{{k_\mathrm{res}}}
\newcommand{\zero}{{_\mathrm{0}}}
\newcommand{\iso}{{_0}}

\newcommand{\s}{\rmn{s}}
\newcommand{\cm}{\rmn{cm}}

\newcommand{\GeV}{\rmn{GeV}}
\newcommand{\pc}{\rmn{pc}}
\newcommand{\kpc}{\rmn{kpc}}
\newcommand{\muG}{\rmn{\mu G}}
\newcommand{\mkG}{\rmn{\mu G}}

\newcommand{\jfm}{J. Fluid Mech.}

\begin{document}

\maketitle

\begin{abstract}
The propagation of charged particles, including cosmic rays,
in a partially ordered magnetic field is characterized by a diffusion tensor
whose components depend on the particle's Larmor radius $\RL$
and the degree of order in the magnetic field.
Most studies of the particle diffusion presuppose a scale separation
between the mean and random magnetic fields
(e.g., there being a pronounced minimum in the magnetic power spectrum
at intermediate scales).
Scale separation is often a good approximation in laboratory plasmas,
but not in most astrophysical environments
such as the interstellar medium (ISM).
Modern simulations of the ISM have numerical resolution of order 1\,pc,
so the Larmor radius of the cosmic rays that dominate in energy density
is at least $10^6$ times smaller than the resolved scales.
Large-scale simulations of cosmic ray propagation in the ISM
thus rely on oversimplified forms of the diffusion tensor.
We take the first steps towards a more realistic description
of cosmic ray diffusion for such simulations,
obtaining direct estimates of the diffusion tensor from test particle
simulations in random magnetic fields
(with the Larmor radius scale being fully resolved),
for a range of particle energies corresponding to
$10^{-2}\lesssim\RL/\lc\lesssim 10^{3}$,
where $\lc$ is the magnetic correlation length.
We obtain explicit expressions for the cosmic ray diffusion tensor
for $\RL/\lc \ll 1$,
that might be used in a sub-grid model of cosmic ray diffusion.
The diffusion coefficients obtained
are closely connected
with existing transport theories that include
the random walk of magnetic lines.
\end{abstract}


\section[]{Introduction}

The theory of propagation and confinement of cosmic rays is largely based on
analytical calculations (mostly quasilinear, applicable if magnetic fluctuations are
much weaker than the background magnetic field) of the scattering of charged relativistic
particles by magnetic inhomogeneities \citep{BerezinskiiEA1990,Sch02,Ku04,SMP07}.
Analytical results have been verified and extended by simulations of the motion of test
particles
\citep{GiacaloneJokipii1994,GiacaloneJokipii1999,CasseEA2001,CandiaRoulet2004,PlotnikovEA2011}
in model magnetic fields
$\vec{B}$ represented as the sum of a mean field $\vec{B}\zero$ and
fluctuations $\vec{b}$,
\begin{equation}\label{BB0b}
\vec{B} = \vec{B}\zero + \vec{b}\,.
\end{equation}
The main goal of the test particle simulations has been the verification of
the analytical results and their extension, especially to the case of finite
magnetic fluctuations \citep[e.g.,][]{SD09}, often focusing on the
dependence of the cosmic ray propagation parameters on the particle energy.

The results are formulated in terms of the cosmic ray diffusion tensor
\begin{equation}
      K_{ij}=K_\perp\delta_{ij}+(K_\parallel-K_\perp)\Bhat_i\Bhat_j\,,
      \label{Difftensor}
\end{equation}
where $\BBhat=\BB/|\BB|$ is the unit vector in the magnetic field direction,
and $K_\parallel$ and $K_\perp$ are, respectively, the cosmic ray diffusivities along and across
the mean magnetic field \citep[e.g.,][and references therein]{GiacaloneJokipii1999}.
Extensive numerical results have been obtained for a wide range of the particle
rigidity, $10^{-2}\lesssim\rho\lesssim10^3$, where
\[
\rho={\RL}/{\lc}\approx{E}/{q}\,,
\]
with
\begin{equation}\label{RL}
\RL\approx3.3\times10^{12}\cm
\left(\frac{B}{1\,\muG}\right)^{-1}
\left(\frac{{E}}{1\,\GeV}\right)\sin\theta
\end{equation}
the Larmor radius of the particle's gyration in the magnetic field (here given for ultra-relativistic
protons), $\lc$ the correlation length of magnetic fluctuations, $E$ and $q$ the total
particle energy and electric charge, and $\theta$ the angle between the magnetic field and the particle
velocity (the pitch angle).
The results of test particle simulations are in
approximate agreement with theoretical estimates and
can be fitted with the form \citep[Section 2.3 of][]{SMP07}
\begin{equation}\label{Kpara}
  K_\parallel = \kappa\zero \left(\frac{E}{1\,\GeV}\right)^n \frac{B_0^2}{b_0^2}\,,
\end{equation}
with an accompanying expression for $K_\parallel/K_\perp$, where $b_0$ is the root-mean-square strength
of magnetic fluctuations, $n=1/3\text{--}1/2$ depending on the magnetic power spectrum,
and $\kappa\zero\simeq(3\text{--}5)\times10^{28}\,\cm^2\,\s^{-1}$ for the Galactic cosmic rays near the Sun.
The spectrum of magnetic fluctuations in the ISM extends from the outer scale of order 100\,pc\footnote{This
estimate of the
largest scale in the interstellar inertial range follows from analyses of the interstellar
velocity field \citep[page 148 in][and references therein]{RSS88}. Observations of interstellar
turbulence in various tracers, e.g., Faraday rotation \citep{MS96} or fluctuations of synchrotron
intensity \citep{HBGM08,Ietal13}, which depend on products of fluctuating quantities, unsurprisingly
suggest smaller values of the turbulent scale as long as they are not interpreted in terms of
the statistical properties of each of the physical variables involved (thermal and cosmic-ray electron
densities, magnetic field, etc.) \citep{SSFTLT14}. Such an interpretation is hampered by the fact
that the variables can be correlated or anticorrelated to a poorly known degree \citep{BSSW03,SSFTLT14}.
}
apparently down to at least $10^6\,\cm$ \citep{ARS95}. Therefore, the Larmor radius of cosmic ray particles
of GeV energies is well within the range of magnetic turbulent scales.

The diffusion of charged particles in a random magnetic field is a result of two dominant
effects. Particles gyrating in a magnetic field are scattered by magnetic fluctuations ---
which may be part of the externally driven magnetic spectrum
or be self-generated by the streaming instability \citep{Ku04} ---
at a scale comparable to $\RL$.
Due to the scattering, the
particle propagation both along and across the magnetic field become diffusive.
The divergence of the field lines of a random magnetic field that guides the particles further
enhances their diffusion. As a result, the propagation of cosmic rays is sensitive to both local and global properties of the magnetic field. \citet{SonsretteeEA2015} present a detailed analysis of the random
walk of magnetic lines with vanishing mean magnetic field \citep[see also][]{SRM13,SonsretteeEA2015b}.

The dominant contribution to the scattering of charged particles is provided by resonant
magnetic fluctuations at a scale close to the particle's Larmor radius given by Equation~\eqref{RL}.
Magnetic fluctuations at
such a small scale can be measured or estimated only in rare
cases, e.g., \textit{in situ\/} in the interplanetary space.
This makes it difficult to apply
the theory of cosmic ray diffusion
to interstellar or intergalactic cosmic rays at their
dominant energies, $E\simeq(1\text{--}10)\times10^3\,\GeV$ per nucleon,
since little beyond crude statistical properties of random magnetic fields
at their outer
(or energy-range) scale is known with any degree of confidence.
This scale is of the order of
$3\times10^{20}\,\cm$ ($100\,\pc$) in the interstellar space and
$3\times10^{22}\,\cm$ ($10\, \kpc$) in galaxy clusters, by far larger than the resonant scale
of the cosmic ray particles.

More importantly, models of cosmic ray propagation rely on the representation \eqref{BB0b}
for magnetic field, which is meaningful and useful only in the case of a clear scale separation
between the mean magnetic field and its fluctuations.
Under such a scale separation, the range of scales (spatial and/or temporal) over which
the mean magnetic field varies is clearly distinct from that of the random magnetic field;
we note that the mean magnetic field does not need to be uniform \citep{GSSFL13b,FS15}.
For example, scale separation can manifest
itself as a local minimum in the magnetic power spectrum at a scale larger than
the outer turbulent scale \citep[see][for a discussion]{GSSFL13b}.
Clear separation of the mean magnetic field and fluctuations is typical of laboratory plasmas
and the solar wind \citep{GRM95,H15},
where the mean and fluctuating magnetic fields vary in space and time at widely separated scales.
However, there are neither theoretical
nor observational reasons to expect that the spectra of interstellar or intergalactic magnetic
fields would have such a feature.
In any event, the existence of such a separation at a scale comparable to a certain value of $\RL$
would justify Equation~\eqref{BB0b} only over a
relatively narrow range of particle energies,
rather than for the wide range of $\RL$ required in a propagation model.
A consistent interpretation of Equation~\eqref{Kpara} could be to use it as
a scale-dependent diffusivity with the scale-dependent `mean' magnetic field
defined as
$B_0^2(k)=\int_0^k  b^2(k')\,\dd k'/k'$
or similarly, but then the propagation of cosmic rays has
to be described with an integral equation which can be reduced, under certain
conditions, to a partial differential equation of a more complicated form than the
diffusion equation
\citep[e.g., Section~5.4 in][see also \citet{SBMS06}]{B08}.

As a result, large-scale simulations of cosmic ray propagation in the galaxies use the forms
\eqref{Difftensor} and \eqref{Kpara}, or even a scalar diffusivity with $\kappa_0$ simply taken
to be constant or a heuristic function of position designed to model galactic spiral arms,
disc and halo.
This is true even of such advanced cosmic ray modelling tools as the \textsc{galprop}
\citep{SM01,SMP07,VDJMMNOPS11},  \textsc{cosmocr} \citep{Miniati01}
and \textsc{dragon} \citep{EGGM08,GMDEG13,Ma13} codes;
see also \citet{Miniati07,HWK09,SOSBH14}.
This is an oversimplification since both parameters that control cosmic ray diffusion,
the ratio of the Larmor radius to the turbulent correlation scale and the strength of magnetic
fluctuations, depend on all MHD variables including gas density, velocity and temperature. As a result,
the relation of the local energy density of cosmic rays to the ISM parameters, a question of primary
importance in radio astronomy, remains uncertain \citep[e.g.,][and references therein]{SSFTLT14}.

One of the goals of this paper is to develop approaches to a more physically detailed modelling of cosmic
ray propagation in numerical magnetohydrodynamic (MHD) simulations of the interstellar medium (ISM). Since the Larmor radius
of cosmic ray particles is by far smaller than the numerical resolution of any ISM simulations,
which tend to resolve scales down to the order of a parsec, a subgrid model of cosmic ray diffusion is required.
Most test-particle simulations of cosmic ray diffusion in the context of the ISM employ a synthetic random magnetic field specified
in continuous space, as in Equation~\eqref{meshless} below
\citep[see, however,][]{CasseEA2001,DeMarcoEA2007,BYL11}, whereas the ISM simulations use a numerical grid with a certain spatial resolution. This should not prevent such results from being used in grid-based simulations, but
we consider test particles motion in both continuous and discrete spatial domains in Section~\ref{MFM}
to appreciate their limitations and select optimal parameter ranges.

Analytical studies of the propagation of charged particles in random magnetic
fields have produced deep insights \citep[see][for a useful review]{CasseEA2001}, many of which
show consistency with test particle simulations. However, such analytical
results are not particularly accurate over a wide range of parameters,
presumably due to the assumptions involved in their derivation.
Here we adopt an empirical approach and aim at deriving approximate asymptotic forms of the diffusion tensor
from test particle simulations, which may be useful in a wide range of applications.
The studies of cosmic ray propagation with test particles often focus on
energetic particles whose Larmor radius is comparable to or exceeds the correlation length
of magnetic fluctuations.
However,  a sub-grid model of cosmic ray diffusion requires the opposite limit,
of a Larmor radius
much smaller than the scales at which magnetic field is known.
This limiting case is our main concern in this paper: we attempt to develop expressions for the cosmic
ray diffusion tensor in a partially ordered, random magnetic field that can be extrapolated to scales
much larger than the particles' Larmor radius.
Thus,
in order to deduce the limiting behaviour, we perform test particle simulations in magnetic fields containing a large
range of scales, but where the particle Larmor radius is
well resolved numerically and is
smaller than
the
correlation length of the magnetic field.
The obtained components of the
cosmic ray diffusion tensor can then be applied to much larger scales, depending on the particle
momentum and the magnetic field strength on those scales.

\section[]{Test particle simulations}

Most test particle simulations employ a synthetic random magnetic field,
either composed of a superposition of static plane waves or specified on a
discrete mesh.
Earlier simulations have demonstrated that the propagation of charged particles
in a random magnetic field is very sensitive to subtle details of the magnetic field structure,
such as the density at which the wave vector space is populated with magnetic modes and
the range of wave numbers involved, etc.\ \citep{CasseEA2001,Parizot2004,FatuzzoEA2010,PlotnikovEA2011}.
Here we describe test particle simulations in which two quite different
models of an isotropic turbulent magnetic field are used and carefully
compared in order to better understand these subtleties. The magnetic fields are taken to be
stationary. We neglect electric fields since they are expected to have a negligible effect on
relativistic particle energy as the acceleration time is likely to be much greater than the
diffusion time \citep{FatuzzoEA2010}.

In one model used here, the
magnetic field is defined at the particle's location using an explicit
algebraic expression. The other
model constructs the magnetic field on a regular mesh,
as in MHD simulations. Equations of motion are then solved for a large
number of particles of a given Larmor radius, over a number of magnetic
field realisations.
The elements of the diffusion tensor are then obtained as functions of time
from the mean-squared particle displacements, with averaging over the particle trajectories
and magnetic field realisations. At a suitably large time,
where the diffusion coefficients have settled to steady values,
asymptotic diffusion coefficients can be obtained.

Our model magnetic field is of the form
\eqref{BB0b}
where
$\vec{B}\zero = B\zero \hat{\vec{z}}$ is a uniform
mean field and $\vec{b}$
is a random component, with $\langle \vec{b} \rangle = \vec{0}$
(where angular brackets denote suitable averaging).
It is convenient to parameterize the turbulence level with
\begin{equation}\label{eta0}
\eta = \frac{b_0^2}{b_0^2 + B_0^2}\,,
\end{equation}
where $0 \le \eta \le 1$ and
$b_0^2=\langle\vec{b}^2\rangle$; we note that $(b_0/B_0)^2=\eta/(1-\eta)$.
Here $\eta=0$ corresponds to a uniform, unperturbed magnetic field, whilst $\eta=1$ implies no mean field,
with the magnetic field being dominated by turbulent small-scale components.
The random magnetic field is assumed to have an isotropic power spectrum
\EQ \label{MS}
M(k)=M\zero
\left\{
\begin{array}{ll}
0               & \mbox{for } k  <   k\zero,  \\
(k/k\zero)^{-s}    & \mbox{for } k \geq k\zero,
\end{array}
\right.
\EN
where $M\zero$ is a normalization constant related to the desired turbulent magnetic energy density $b_0^2$,
and $k\zero = 2 \pi / L$,  with $L$ the largest (outer) scale of turbulence.
Therefore, the spectral power of the magnetic field is
localised at well-separated wave numbers, $k\geq k_0$
due to the random field, and $k=0$ due to the mean (in this case, uniform) part.

We consider two values of the spectral index, $s = 5/3$ or $3/2$, to explore
its effects on the particle transport.
The correlation length $\lc$ of an isotropic field is given by \citep[][]{MoninYaglom1975}
\EQ \label{eq:lc}
\lc = \frac{\pi}{2}\frac{\int_0^{\infty}k^{-1}M(k) \,\dd k}{\int_0^{\infty}M(k) \,\dd k}\,,
\EN
with $\lc =0.1 L = \pi / (5 k\zero)$ for $s=5/3$, and $\lc= \pi / (6 k\zero)$ for $s=3/2$.
Our numerical implementations of the random magnetic field have discrete values of the wave
number $k$ and extend to some maximum $k = \kmax$. As a result, the effective
value of the correlation length differs by up to a few per cent from the above
values obtained for $\kmax\to\infty$.
We also considered a magnetic spectrum of the form
\begin{equation}\label{Mkb}
M(k)=M_0 \frac{(k/k_b)^4}{[1 + (k/k_b)^2]^{s/2 + 2}}\,,
\end{equation}
which is similar to Equation~(\ref{MS}), but has
the peak in the energy spectrum near some chosen $k_b$ rather than at $k\zero$.

\subsection[]{Magnetic field models}\label{MFM}

The first approach is to represent
the magnetic field as a sum of a finite number of plane-wave modes with random
polarizations, wave vectors and phases. It can be shown
\citep[e.g.,][]{Batchelor1953} that the resulting field is spatially homogeneous and
isotropic in the limit of an infinite number of modes.
\citet{GiacaloneJokipii1994} implemented this idea which
has subsequently been used in many studies involving
test
particle simulations.
Here we use a very similar method based on a time-independent version
of the velocity field
used by \citet{FungEA1992} to model synthetic turbulence (and
often used in modelling of turbulent flows
by other authors). We take
\EQ \label{meshless}
\vec{b}(\vec{x}) = \sum_{n=1}^{N}
	\left[\vec{C}_{n} \cos (\vec{k}_n \cdot \vec{x}) + \vec{D}_{n} \sin (\vec{k}_n \cdot \vec{x})\right],
\EN
where $\vec{k}_n$ are randomly oriented wave vectors, and $\vec{C}_n$ and $\vec{D}_n$ are
random vectors confined to the plane perpendicular to $\vec{k}_n$ to
ensure that $\nabla \cdot \vec{b} =0$.
With
\EQ\label{CDn}
|\vec{C}_n|^2=|\vec{D}_n|^2= \frac{\Delta k_n M(k_n)}{\sum_{n=1}^N \Delta k_n M(k_n)},
\EN
where $\Delta k_n$ is a suitable wave number spacing as defined below, the
desired energy spectrum is obtained. The random angle between the amplitude vectors controls the random polarization of the plane waves.
An efficient way to model a power-law spectrum is to select geometrically spaced wave numbers $k_n$
between $k\zero$ and $\kmax$:
\EQ
k_n = k\zero \left(\frac{k_{N}}{k\zero}\right)^{(n-1)/(N - 1)},
\EN
where $1 \le n \le N$ (so that $k_1 = k\zero$ and $k_{N} = \kmax$).
In Equation~\eqref{CDn}, we then take
$\Delta k_1 = (k_2 - k_1)$, $\Delta k_N = (k_N - k_{N-1})$,
and otherwise $\Delta k_n=(k_{n+1} - k_{n-1})/2$, for $2\le n \le N-1$.
For large $N$, the resulting magnetic field
is essentially non-periodic, which is potentially useful in detecting periodicity effects when compared with the second (periodic) model described below. An appropriate value for $N$
depends on the context; we examine this in Appendix~\ref{results:comparison}.
This model of a random magnetic field will be henceforth referred to as the
\textit{continuum model}.

The second model implements
a spatially periodic random magnetic field on
a regular, three-dimensional Cartesian mesh, similarly to  \citet{CasseEA2001}.
First, a regular grid in the three-dimensional wave number space is
populated with complex Fourier components $\vec{b}(\vec{k})$ given by
\EQ \label{bkmesh}
\vec{b}(\vec{k}) = A(k)[ \vec{b}_1(\vec{k}) e^{{\rm i}\phi_1(\vec{k})} +
{\rm i}\vec{b}_2(\vec{k}) e^{{\rm i} \phi_2(\vec{k})} ],
\EN
where $\vec{b}_1$ and $\vec{b}_2$ are random, real unit vectors perpendicular to
both $\vec{k}$ (to ensure that $\nabla \cdot \vec{b}=0$) and each other,
so that $\vec{b}_1$ is chosen at random and $\vec{b}_2=\widehat{\vec{k} \times \vec{b}_1}$, where the hat denotes a unit vector.
Here $\phi_1(\vec{k})$ and $\phi_2(\vec{k})$ are distinct random phases that control the polarization
of the magnetic field modes.
We note that results presented below, at least for $\eta=1$, are not particularly sensitive to the
random phases, and taking $\phi_1(\vec{k})=\phi_2(\vec{k}) = 0$ changes the results very little:
the randomness in the choice of $\vec{b}_1$ is sufficient to ensure randomness in the
relative phases of different modes.
We take $A(k) \propto \sqrt{M(k)}/k$ to set the correct
energy spectrum. We enforce Hermitian symmetry on $\vec{b}(\vec{k})$ so
that the inverse Fourier transform of Equation~\eqref{bkmesh} yields a real,
random magnetic field which we then normalise to the desired energy density
$b_0^2$.
In what follows, this model is referred to as the
\textit{discrete model}.
The appropriate grid size is discussed in Appendix~\ref{results:comparison} where
the two magnetic field models are compared.
We select a sufficiently small grid spacing in the discrete model,
and a sufficiently large maximum wave number $\kmax$ in the continuum model, to ensure that
the magnetic field is fully resolved at the Larmor radius scale.

\label{subtleties}There are a few important subtleties in the numerical implementation of random magnetic fields on a mesh
(i.e.\ in the discrete model),
arising from the requirements of isotropy and randomness of phases (to a given accuracy).
To ensure these at the outer scale $k\zero$, there should be a sufficiently large number of modes
at this wave number. In other words, there should be a sufficiently large sphere of (approximately) this radius
in $k$-space. Failing to do this may affect the particle diffusion significantly.
Therefore, the size of the computational domain in $k$-space should be larger
than $\kmax/k\zero$, and, correspondingly, the domain in physical space should be larger than $L=2\pi/k\zero$.
If the isotropy and the randomness of phases have been achieved at the outer scale, smaller scales
do not represent a problem in this respect.
In addition to magnetic modes at
the resonant wave number $k \approx 2\pi/\RL$, previous studies have shown that
the density of nearby wave numbers can be important for effective
particle scattering with some types of magnetic field \citep[e.g.,][]{Mace12}.
Similar consideration may be important for both of our magnetic field
constructions.
We carefully test our numerical implementations of the magnetic field for isotropy
at all relevant scales including $2\pi/k\zero$ to ensure that the modes of a given wave number
have randomly distributed phases.
It seems that taking a mesh size
  ${N_x = N_y = N_z} \ge 4 \kmax/k\zero$ is sufficient if $\kmax$ is set to be the radius of the largest sphere that fits within the wave number domain. Larger mesh sizes may be preferable when $R\larmor \gtrsim L$, but in that case one might compensate by reducing $\kmax/k\zero$, since the small scales should have little influence on
the particle transport \citep{PlotnikovEA2011}.

We use a trilinear interpolation from the nearest grid points to obtain
the magnetic field between them.
On one server with 256\,GB
of RAM we could achieve a resolution of up to
$2048^3$ mesh points. Larger resolutions, involving a mesh
distributed in memory across multiple compute nodes, are much more computationally
expensive and are not likely to provide any additional benefit in the present study. A
magnetic field realization can be computed in much less than a minute of CPU time at a resolution $1024^3$.
It is desirable to use a large number of the magnetic field realisations, with as few
particles per realisation as possible. For example, at $1024^3$ around $30$
realizations, each with about $50\text{--}100$ particles can be performed within a reasonable time.
However, in such a time, more particles per realisation have to be used
as the mesh size increases in order to obtain sufficient statistics.
Once the magnetic field has been computed, solving the particle equations of
motion is relatively quick.

When magnetic field is defined using Equation~\eqref{meshless}
(i.e.\ in the continuum model),
very little
computer memory is needed because the field is computed only at the location of a
particle. In this case, the number of modes $N$ is limited by the computer time available
because the field is computed repeatedly at each time step when the equations of motion are
solved.

\subsection[]{Equations of test particle motion}

The position
$\vec{x}$
of an energetic charged particle
in magnetic field $\vec{B}$
is governed by the Newton--Lorentz
equations,
\EQ\label{eqnmot}
\frac{d \vec{x}}{dt} = \vec{v}\,,
\qquad
\frac{d \vec{v}}{dt} = \frac{q}{\gamma m c}\vec{v} \times \vec{B}\,,
\EN
where $\vec{v}$ is the particle velocity,
$q$ and $m$ are the particle charge and rest mass,
$\gamma =\sqrt{1-v^2/c^2}$ is the Lorentz factor, $v=|\vec{v}|$, and $c$ is the speed of light.
Since the electric field has been neglected, particle energy is conserved.
It is useful to define dimensionless variables
\[
\vec{x}' = \frac{\vec{x}}{L}\,,
\quad
\vec{v}' = \frac{\vec{v}}{v\zero}\,,
\quad
\vec{B}' = \frac{\vec{B}}{\Brms}\,,
\quad
t' = \frac{t}{t\zero}\,,
\]
with $t\zero=L/v\zero$ and $\Brms=(B_0^2+b_0^2)^{1/2}$,
in terms of $L$ and an appropriate reference
speed $v\zero$;
for relativistic particles, $v\zero\approx c$ is an appropriate choice.
The resulting dimensionless form of Equations~\eqref{eqnmot} is given by
\EQ\label{eqnmotp}
\frac{d \vec{x}'}{dt'} = \vec{v}'\,,
\qquad
\frac{d \vec{v}'}{dt'} = \alpha \vec{v}' \times \vec{B}'\,,
\EN
where
\EQ
\alpha = \frac{q  L\Brms}{\gamma m cv\zero}
\, .
\EN
In this work we examine how the particle transport depends on the
characteristic Larmor radius,
\EQ
\RL =  \frac{\gamma m v c}{|q|\Brms} = \frac{pc}{|q|\Brms} = \frac{pc}{|q|}\frac{\sqrt{\eta}}{b_0}\,,
\EN
where $p=\gamma m v$ is the particle momentum.
We note for the sake of the reader's convenience that $\RL$ is related to
the particle kinetic energy via $E_k = \sqrt{(\RL q \Brms)^2 + m^2c^4} - mc^2$,
or $E_k \propto \RL$ for relativistic protons in a microgauss-strength
magnetic field.
The dimensionless Larmor radius is given by
\[
R_\mathrm{L}' = \RL / L = v' /\alpha\,.
\]
We note that the Larmor radius
is defined here with respect to the total rather than mean magnetic field strength.

Individual particles have random initial positions within a
cube of approximate length $L$, the outer scale
of the turbulence.
The particle velocity is given a random initial direction with $|\vec{v}'| = 1$.
We then solve equations of motion numerically for $\Brms'=1$ and set
$R_\mathrm{L}'=\alpha^{-1}$ by selecting $\alpha$ as desired.

We solve Equations~\eqref{eqnmotp} numerically, primarily using the fifth-order
method of \citet{CashKarp90}, which allows for an adaptive step size via the
error estimate of the fourth-order solution. Additionally, in order to check if precise
energy conservation is essential,
we have sometimes used the highly conservative
method of \cite{Boris70}, which is widely used in particle-in-cell (PIC) plasma simulations.
We find no appreciable difference in results between the two methods; however, the former is more
efficient. Over all results reported here, the maximum particle kinetic energy change
over the diffusion time $t_{\mathrm{max}}$ -- which is typically $(10^1\text{--}10^2) t\zero$, as is seen in Figure~\ref{evo} --
is around $0.1$ per cent (but is often much less).

\subsection[]{Diffusion coefficients}

Time-dependent diffusion coefficients were obtained from test particle
simulations using
\EQ \label{Running}
\kappa_{xx}(t) = \frac{1}{2}\frac{\dd} {\dd t}\langle (\Delta x )^2 \rangle=
\langle v_x(t) \Delta x \rangle\,,
\EN
where $\Delta x$ is the particle displacement in the $x$-direction over some
time $t$, and $v_x(t)$ is the $x$-component of the particle velocity at the end of the displacement.
The angular brackets in Equation~\eqref{Running}
denote averaging over displacements for multiple time intervals of length
$t$
on a specific particle trajectory,
over many particles and several (ideally, equally many) magnetic field realisations.
Similar expressions were used to evaluate
$\kappa_{yy}(t)$ and $\kappa_{zz}(t)$.
In this work we are mainly interested in the asymptotic limit,
$\kappa_{ij} = \lim_{t \to \infty} \kappa_{ij}(t)$, which can be approximated by taking
$\kappa_{ij}(t)$ at a suitably large time.
In this limit, Equation~(\ref{Running}) is equivalent to
\EQ \label{Partial}
\kappa_{xx} = \lim_{t \to \infty}\frac{\langle ( \Delta x )^2 \rangle}{2 t},
\EN
which has been employed in other works \citep[Section 1.3.1 of][and references therein]{Sh09}.
As illustrated and discussed in Appendix~\ref{AppA}, the two prescriptions are equivalent but
the former is more convenient in simulations, especially when the time dependence of
the diffusivity is as smooth and simple as in our case. In particular,
the asymptotic diffusion coefficients are obtained from Equation~\eqref{Running} at
an earlier computational time than with Equation~(\ref{Partial}).

Here we have no real interest in the off-diagonal elements $\kappa_{ij}$, where $i \neq j$,
since they describe systematic drifts of cosmic-ray particles
that are not likely to affect their large-scale diffusion \citep[Section 1.6 of][]{Sh09}.
We use the difference between the numerically obtained $\kappa_{xx}(t)$ and $\kappa_{yy}(t)$ as
an estimate of the error in the diffusion coefficients; typically we obtain sufficient statistics to have the maximum running difference of
less than three per cent.
Note that for the diagonal elements of $\kappa_{ij}$
--- but not, in general, for the off-diagonal ones \citep{Sh11} ---
the above relations produce
the same result as the standard
{Taylor--Green--Kubo}
formula \citep{Taylor22,Green51,Kubo57}
\EQ \label{TGK}
\kappa_{ij}(t) = \int_0^{\infty} \langle v_i(0)v_j(t)\rangle dt,
\EN
where $v_i(t)$ is the velocity component in the direction $x_i$ at time $t$.

The diffusion coefficients $\kappa_{ij}$ thus obtained are related, via a relation similar
to Equation~\eqref{Difftensor}, to the diffusivities $\kappa_\perp$ and $\kappa_\parallel$
across and along the magnetic field at the largest scale available, which is
comparable to $L$ in our implementations of the random magnetic field. In contrast, the parallel
and perpendicular diffusivities $K_\parallel$ and $K_\perp$ of Equation~\eqref{Difftensor} are
defined with respect to $\vec{B}_0$, the magnetic field averaged over
scales larger than the turbulent ones. Therefore, we call $\kappa_{ij}$, $\kappa_\parallel$ and
$\kappa_\perp$ the \textit{local} diffusivities. In the context of sub-grid modelling of cosmic ray
diffusion, the scale $L$ is close to the smallest resolved scale in an MHD simulation,
and the effects of magnetic fields at larger scales,
approximated by Equation~\eqref{Difftensor},
should be faithfully reproduced by the cosmic-ray propagation equation
with only the resolved magnetic field,
using the diffusion tensor $\kappa_{ij}$.

The continuum and discrete magnetic field models are compared in Appendix~\ref{AppA} where we
present the running diffusion coefficient obtained for a various values of $\RL/\lc$ and numerical
resolutions. The range of variation in the values of the diffusivity,
obtained under widely varying numerical resolutions ($512^3$--$1536^3$)
in model \eqref{bkmesh},
and number of modes ($N=150$--$1024$) in model \eqref{meshless},
decreases from less than 20 per cent
for $\RL/\lc=5$ to about 5 per cent for $\RL/\lc\leq1$.
Based on these tests, we initially decided to use
$N=150$ and $\kmax/k\zero = 96$ with the continuous model \eqref{meshless}
and the resolution $512^3$ with $\kmax/k\zero=128$ with the discrete model \eqref{bkmesh}.

\section[]{Diffusion in a purely random magnetic field}
\label{section:zeromean}
\begin{figure}
  \begin{center}
    \includegraphics[width=0.45\textwidth]{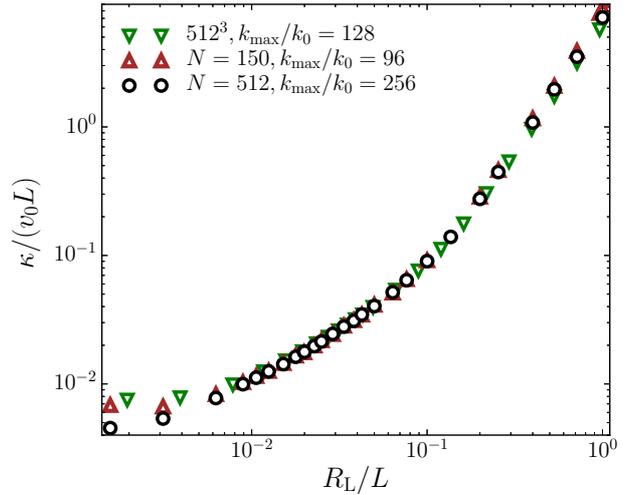}
  \end{center}
  \caption{\label{fig:asymp} Different implementations
  of a purely random magnetic field (with $s=5/3$) lead to consistent values of the
  isotropic diffusivity for $\RL/\lc\ll1$ (with $\lc \approx 0.1 L$). Circles and upward triangles are for the
  continuous model, and downward triangles are for the discrete model.
For $\RL > \lc$, the diffusivity scales as $\RL^2$, but the results
   from the different magnetic field models diverge with increasing $\RL$ for the reasons
   discussed in Appendix~\ref{results:comparison}. For $s=3/2$, similar trends
   are observed.}
\end{figure}

\begin{figure}
  \begin{center}
    \includegraphics[width=0.45\textwidth]{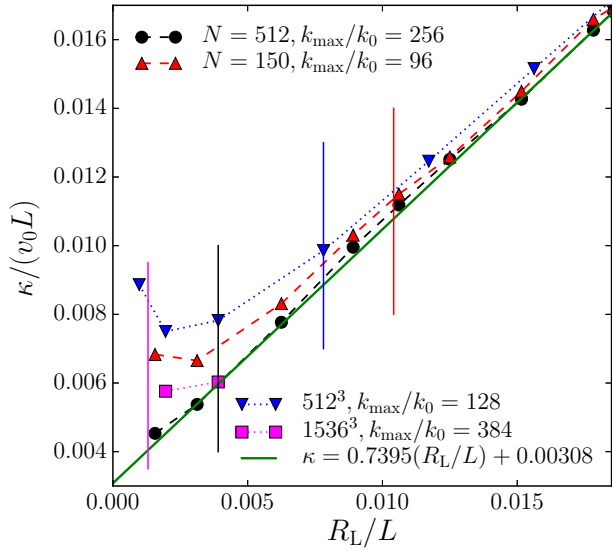}
  \end{center}
  \caption{\label{fig:res} The isotropic diffusion coefficient $\kappa_0$
    at $\RL \ll \lc \approx 0.1L$ for $\eta=1$ and $s=5/3$ with the best fit to the most highly-resolved data
  shown with (green) solid line   (see the text and Figure~\ref{fig:fit}).
  The linear dependence of $\kappa_0$ on $\RL$ extends to very small values of $\RL$,
  provided $\kmax/k\zero$ is large enough. The vertical lines of the corresponding colour
  indicate $\RL=2\pi/\kmax$ for the various parameter values used. The numerical resolutions,
  mode numbers and magnetic field models used are specified in the legend. Deviations
  from the asymptotic linear dependence of $\kappa_0$ on $\RL/L$ at small values of this ratio
  are sensitive to these factors.}
\end{figure}

Figure~\ref{fig:asymp} shows the asymptotic (in time) isotropic diffusion coefficient $\kappa_0$
as a function of $\RL/L$ for $s=5/3$ and $\eta=1$ (zero mean field) obtained with various magnetic field
models and resolutions. A linear scaling at $\RL \ll \lc$ and a quadratic scaling at $\RL \gg \lc$
are found.
The latter scaling is consistent with both theory and earlier test particle simulations
\citep[see][and references therein]{Parizot2004,PlotnikovEA2011}.
However, the differences between the diffusivities obtained with different magnetic
field models increase with $\RL/L$  for a \textit{fixed} resolution and $\kmax/k\zero$.
As discussed in Appendix~\ref{results:comparison}, this can be remedied by
changing the range, number density and/or location of the magnetic modes in $k$-space.

On the other hand, results at $\RL\ll\lc$ exhibit an encouragingly weak dependence on the
details of the magnetic field model. Figure~\ref{fig:res} illustrates this regime
in finer detail where we have added additional results from a discrete magnetic field
model at a resolution $1536^3$ with $\kmax$ three times larger than that with the $512^3$
resolution. In the $512^3$ case, deviations from the linear trend shown by the solid line
line start almost from the extreme right of the
range of $\RL$ shown. The solid blue vertical line (the second vertical line
from the right) shows where $\RL = 2\pi/\kmax$, i.e.\ at the scale of the
smallest structures in the magnetic field; significant deviations from the linear
trend occur below this marker.
With the $1536^3$ grid, deviations from the straight line are much weaker (the leftmost
vertical line marks $\RL = 2\pi/\kmax$ for this case). With the continuum magnetic
field model, the deviation from the linear trend is weaker than that in the case of a discrete model
with a similar value of $\kmax$.
These results seem to indicate that the continuation of the linear trend
down to smaller $\RL$ depends on the presence of magnetic modes at or near to the resonant wave number
$\kres=2\pi/\RL$.
Suppression of such modes presumably
leads to an increased particle mean free path parallel
  to the local magnetic field, resulting in an enhanced diffusion coefficient, as observed in
  Figure~\ref{fig:res}. Given that the continuum model has very few modes at large $k$, it is perhaps
  surprising that deviation from the linear trend occurs at approximately $\RL=2\pi/\kmax$, whereas the
  discrete magnetic field model, containing many more modes near that scale, deviates from the linear trend
  at a somewhat higher value of $\RL$. One reason for this can be that
  the
  high-$k$ modes in the discrete model are not sufficiently well resolved in
  $k$-space.
It may also
  be the case that the numerical construction curtails energy at the smallest scales.
  Furthermore, at these extremes $\RL$ is close to the grid resolution on which the magnetic field is
  defined, and errors from the trilinear interpolation between mesh points can have some
effect.
These results suggest that caution is required, and that $\RL$ should
not be less than about three separations of the mesh on which magnetic field is defined.

Simulations for still smaller values of $\RL/L$ are more computationally efficient with the
continuous magnetic field model. We explored the range down to $\RL/L=10^{-3}$ using up to
$N=5000$ modes; the diffusion coefficient remains very close to the
solid line shown in Figure~\ref{fig:res}. \citet{CasseEA2001} found a Bohm scaling
of the diffusion coefficient, $\kappa \propto \RL v\zero$, similar to our results
in a limited range of $\RL$. However, these authors find a different
scaling below a certain value of $\RL$ \citep[see also][]{FatuzzoEA2010}.
It is not clear why the result found here is different
from those; the most obvious difference is that those simulations used
$\kmax/k\zero=10^4$ with only $N=200$ modes in their continuous magnetic field model. Here we
confirm the linear scaling of $\kappa_0$ with $\RL/\lc$ down to very small values
of $\RL/\lc$ with both magnetic field implementations.

\begin{figure}
  \begin{center}
    \includegraphics[width=0.45\textwidth]{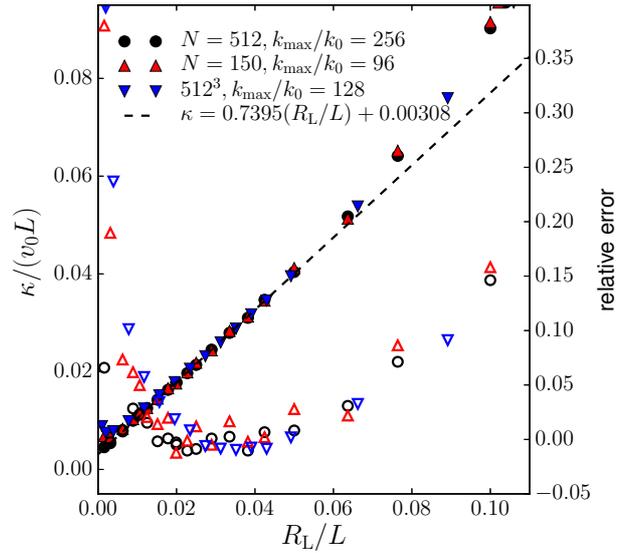}
  \end{center}
  \caption{\label{fig:fit} The isotropic diffusivity $\kappa_0$ at low $\RL$
  in a continuous and discrete random magnetic fields with the number of modes $N$,
  $\kmax/k\zero$ and the resolution specified in the legend. The best fit
  of Equation~\eqref{eq:kappaiso} is shown with dashed line (also shown solid in
  Figure~\ref{fig:res}). Open symbols show the corresponding relative deviations from the
  best fit (the right-hand ordinate axis).}
\end{figure}

In Figure~\ref{fig:fit} we show a fit of the form
\EQ \label{eq:kappaiso}
\frac{\kappa\iso(\RL)}{v\zero L} \approx 0.0031 + 0.74 \, \frac{\RL}{L}
\EN
 to the isotropic diffusivity obtained with $\kmax/k\zero=256$
over the range where the variation with $\RL/L$ is linear, $4\times10^{-3}\lesssim\RL/L\lesssim5\times10^{-2}$
with $\lc \approx 0.1$.
The relative deviations from the fit are shown in Figure~\ref{fig:asymp}; they
do not exceed about 15 per cent at $\RL = \lc$ (i.e., $\RL/L\approx0.1$), where we might have expected
more deviation, given the tendency towards quadratic scaling with $\RL$ at around $\RL=l_c$.
The fit approximates the numerical results to within 3 per cent for all values of $\RL$,
between $\RL=\lc/2$ and the smallest values of $\RL$ considered (where already $\RL < 2\pi/\kmax$,
resulting in reduced resonant scattering).

The fit has a non-zero intercept,
\begin{equation}\label{intercept}
\left.{\kappa\iso}\right|_{\RL\to 0}\approx 3\times10^{-3}v\zero L\,,
\end{equation}
corresponding to a part of the diffusivity independent of the Larmor radius (i.e., the particle energy).
For $v\zero=c$, and $L\simeq100\,\pc$ for the outer turbulent scale, we obtain
$\kappa\iso \simeq 3 \times 10^{28}\, \cm^2\, s^{-1}$.
We can compare this result to an estimate based on the particle diffusion along random magnetic lines \citep[][]{JokipiiParker1968}, known as the field line random walk (FLRW) model.
If particles are tied to the magnetic lines and maintain a constant pitch angle,
the diffusion coefficient perpendicular to the magnetic field can be estimated as the magnetic field
line diffusion coefficient
(defined as the spatial divergence rate of the magnetic lines -- see below)
multiplied by the particle velocity parallel to the
field lines, i.e., $\kappa =
{\tilde{v}} \Db /3$, where $\Db$ is the magnetic field line
diffusion coefficient, as defined below, and
${\tilde{v}}/3$ is the characteristic
particle speed along the field line, with the factor $1/3$ being appropriate for
the isotropic case. Note that it is more intuitive to make
such a calculation when $\eta \ll 1$, where the diffusion coefficient is
essentially measured perpendicular to the mean
magnetic field line; nevertheless,
here we will associate $\kappa$ with perpendicular diffusion, although there is
no preferred direction in the isotropic case $\eta=1$.
Let us now define $\Db$ for a general $\eta$, since it will be useful in
subsequent sections where $\eta < 1$.
Consider random magnetic field line displacements $\Delta x_{i}^{\mathrm{FL}}$ as a function of the
arc length $s$, measured along the magnetic line (i.e. the change in field line
position with $s$).
Then defining
\EQ \label{eq:D}
D_{b,i} = \tfrac12 \lim_{s \to \infty} \frac{\dd \langle (\Delta x_i^{\mathrm{FL}})^2 \rangle}{\dd s}\,,
\EN
we take $\Db = \tfrac13\sum_{i=1}^3 D_{b,i}$ for $\eta=1$ and
$\Db=\tfrac12\sum_{i=1}^2 D_{b,i}$ when $\eta < 1$ and $\vec{B}_0\parallel\widehat{\vec{x}}_3$.
Note that $\Db$ has the dimension of
length.
We evaluate Equation~\eqref{eq:D} numerically by solving field line equations for many field lines
over several magnetic field realisations, essentially as described by \cite{SonsretteeEA2015}.
Alternatively, \citet{SonsretteeEA2015} provide an analytical approximation of $\Db$ in terms of
$\lc$ or another length scale from the magnetic power spectrum. In both cases, for $\eta=1$ we find that
$\Db \approx 0.06 L$. Due to the pitch angle scattering, particles travel back
and forth along magnetic lines, so that
${\tilde{v}}$ can be much smaller than the
instantaneous particle speed $v\zero$. Equation~\eqref{intercept} is obtained
for ${\tilde{v}} \approx 0.15v\zero$. If this model is correct, then
the parallel scattering would have to be non-diffusive, otherwise we would obtain
compound subdiffusion \citep[][]{Urch1977,KotaJokipii2000} perpendicular to the magnetic lines. A more likely scenario is
that particles eventually escape to nearby field lines, so that the FLRW limit is invalid.

To gain further insight into the nature of Equation~\eqref{eq:kappaiso},
we considered various forms of the magnetic power spectrum $M(k)$.
When its slope is $s=3/2$, rather than $s=5/3$, we have
\[ 
\frac{\kappa\iso(\RL)}{v\zero L}\approx 0.0019  + 0.76 \,\frac{\RL}{L} \,,
\]
that is, a similar slope with the intercept different by 30 per cent from that in
Equation~(\ref{eq:kappaiso}).
We also considered magnetic spectrum of the form \eqref{Mkb} with $s=5/3$ and $k_b=5k\zero$
to obtain
\[
\frac{\kappa\iso(\RL)}{v\zero L}\approx 0.0017 + 0.75\,\frac{\RL}{L} \, ,
\]
which, again, is similar to the other two forms but with yet another value for the
intercept, about a half of that obtained
for $s=5/3$.
This suggests that $\left.{\kappa\iso}\right|_{\RL\to0}$
depends on the form of the magnetic power spectrum,
but that this dependence is not very strong,
variations remaining within a factor of two
for reasonable forms of the spectrum.

\section[]{Diffusion in a partially ordered magnetic field}
In this section we consider cosmic ray diffusion in a magnetic field with  $\eta<1$,
where $\vec{B}_0$ is aligned with the $z$-axis
($z\equiv x_3$).

\begin{figure}
  \begin{center}
    \includegraphics[width=0.45\textwidth]{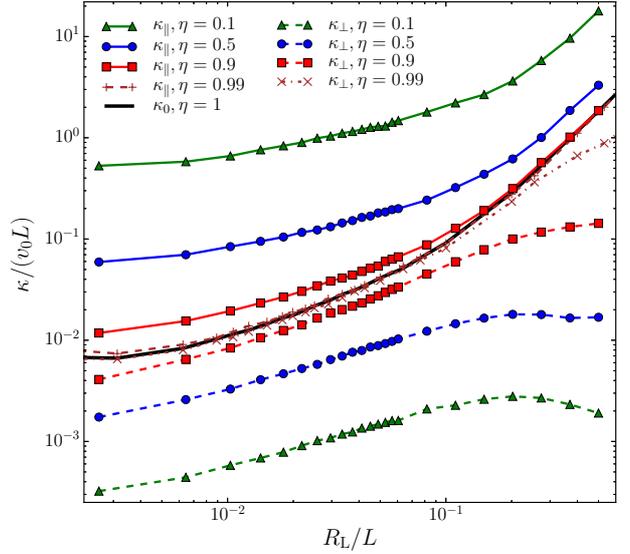}
  \end{center}
  \caption{\label{fig:nonzero} Asymptotic (in time) diffusion coefficients
  $\kappa_\perp$ and $\kappa_\parallel$, for various
    values of $\eta$ specified in the legend, as functions of $\RL$ for the continuous
    magnetic field model with $s=5/3$ and $\kmax/k\zero = 200$. The black solid curve
    shows the diffusivity
    in the corresponding purely random magnetic field as in Section~\ref{section:zeromean}.
    }
\end{figure}

Figure~\ref{fig:nonzero} shows the \textit{local} parallel and perpendicular diffusion
coefficients $\kappa_\parallel$ and $\kappa_\perp$ for several values of $\eta$ as a
function of $\RL$, for the magnetic spectrum \eqref{MS} with $s=5/3$. A reduction in
the pitch angle scattering is expected where
$\RL<2\pi/\kmax$, that is $\RL/L < k\zero/\kmax = 5\times10^{-3}$ (the latter equality applies
for the parameter values used in Figure~\ref{fig:nonzero}). This is
  evidenced in the deviation from the trend of the diffusion coefficients
  at the smallest values of $\RL/L$ plotted in the figure. Although this is not
  easy to see in the figure itself, the behaviour is similar to that seen in
Figure~\ref{fig:res} when plotted on a linear scale.
For a given $\RL$, the diffusivity in a purely random magnetic field, $\kappa\iso$ obtained at $\eta=1$,
separates regions in Figure~\ref{fig:nonzero} filled with the curves representing
$\kappa_\parallel$ in the upper part and $\kappa_\perp$ in the lower:
$\kappa_\perp<\kappa\iso<\kappa_\parallel$.

Before considering the range $\RL/L\ll1$, our main concern here, a few comments
on the region of larger values of $\RL$, where $\RL/L \ge \lc/L=0.1$ might be appropriate.
Here $\kappa_{\parallel} \propto \RL^2$, while $\kappa_{\perp}$
tends to become independent of $\RL/L$. \citet{PlotnikovEA2011} explored
the range $\eta \ge 0.5$ to find that $\kappa_{\perp}/\kappa_{\parallel}\propto\RL^{-2}$
at large $\RL$, even when $\eta$ is very close to unity
(which implies a significant reduction of $\kappa_\perp$ from $\kappa\iso$
for any $B\zero>0$). Our results reproduce this behaviour for $\eta=0.5$,
but we have not explored large enough $\RL$ to observe it at larger values of $\eta$.
However, $\kappa_{\perp}$ decreases with increasing $\RL$ for $\eta=0.1$ and 0.5,
rather than becoming independent of $\RL/L$. In the rest of this section we consider small $\RL$.

\subsection[]{Parallel diffusion}
\begin{figure}
  \begin{center}
    \includegraphics[width=0.45\textwidth]{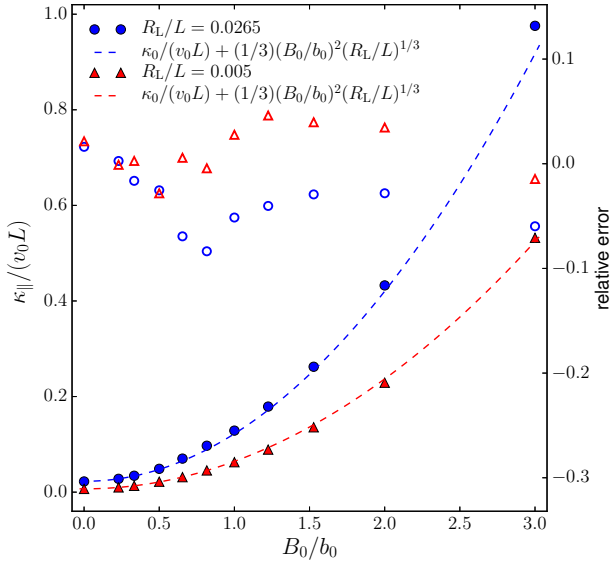}
  \end{center}
  \caption{\label{fig:kpara1} The local parallel diffusivity $\kappa_{\parallel}$
  plotted versus $B\zero/b_0=\sqrt{(1-\eta)/\eta}$ for two values of $\RL$,
  together with their approximations (see the text).
Filled (blue) circles are for $\RL = \lc/4$, where the relative error in the approximation
is less than 10 per cent for the range of $B\zero/b_0$ shown ($0.1 \le \eta \le 1$)
(as shown by open circles, with the axis on the right of the frame).
For the smaller value $\RL/L=0.005$ (filled triangles), the error
(open triangles) is still smaller at a few per cent.}
\end{figure}
\begin{figure}
  \begin{center}
    \includegraphics[width=0.45\textwidth]{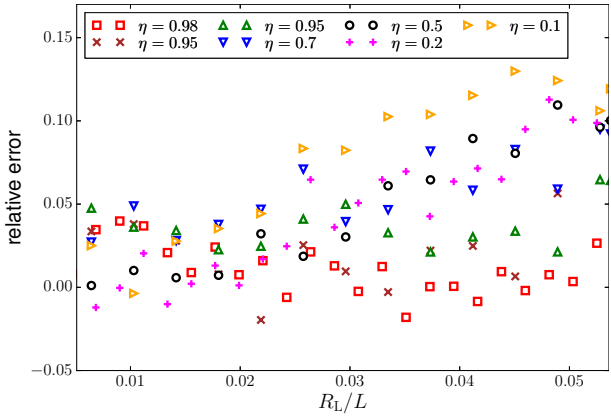}
  \end{center}
  \caption{\label{fig:kparaerror} The relative error in the approximation
for $\kappa_{\parallel}$ given by Equation~(\ref{eq:kpara}) to the test particle
simulations for various values of $\eta$ given in the legend. The
largest values of $\RL$ shown correspond to $\RL \approx \lc/2$.}
\end{figure}

For a given $\RL$, we find that $\kappa_{\parallel}$ scales quadratically
with $B\zero/b_0$ --- or linearly with $(1-\eta)/\eta$ --- as shown in Figure~\ref{fig:kpara1}.
The numerical results can be represented by $\kappa\iso$ of Equation~(\ref{eq:kappaiso})
with the addition of a slowly varying power law function of $\RL$,
of the form $aR_\mathrm{L}^b$. We fitted the two
parameters $a$ and $b$ at small values of $\RL$ to obtain the following approximation
shown in Figure~\ref{fig:kpara1} with dashed lines:
\EQ \label{eq:kpara}
\frac{\kappa_{\parallel}(\RL,\eta)}{v\zero L} \approx \frac{\kappa\iso(\RL)}{v\zero L} + 
\tfrac{1}{3} \left(\frac{\RL}{L}\right)^{1/3} \frac{1-\eta}{\eta}\,.
\EN
We note that the final term has the same scaling as is obtained from quasilinear theory \citep{Jokipii1966}.
  In particular, for the magnetic spectrum~\eqref{MS}, and assuming magnetohydrodynamic waves propagating parallel to the mean field, the appropriate quasilinear result is
\citep[e.g.,][and references therein]{BerezinskiiEA1990,Sch02}
\EQ \label{eq:kparaQL}
\kappa_{\parallel}^{\mathrm QL} = \frac{4 v\zero L \left(2\pi\right)^{-s} }{(s-1)(2-s)(4-s)}\left(\frac{B\zero}{b\zero}\right)^2 \left(\frac{\RL}{L}\right)^{2-s},
\EN
which for $s=5/3$ gives
\EQ
\frac{\kappa_{\parallel}^{\mathrm QL}}{v\zero L} \approx 0.36\left(\frac{\RL}{L}\right)^{1/3} \frac{1-\eta}{\eta}\,,
\EN
which is in
very good agreement with the fitted term above.
For
$s=3/2$ and $\eta=0.5$ a fit to the data yields a similar scaling.
An important point to note is that the usual definition of $\RL$ in Equation~\eqref{eq:kparaQL} (i.e., with respect to the mean field) deviates from
our definition in the limit $\eta \to 1$.
Figure~\ref{fig:kparaerror} demonstrates the fair quality of approximation of the test particle
simulations for various $\eta$ and $\RL$.
For $\RL \le \lc/4 \approx 0.025L$, the accuracy is better than five per cent, and
within 10 per cent at $\RL = \lc/2$.

\subsection[]{Perpendicular diffusion for $\RL \ll \lc$}\label{perpdiff}
\begin{figure}
  \begin{center}
    \includegraphics[width=0.45\textwidth]{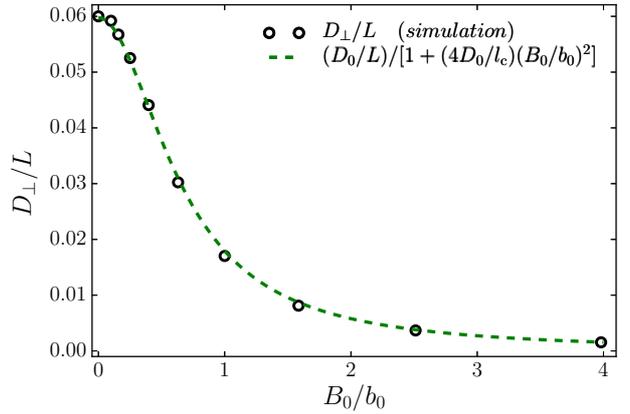}
  \end{center}
  \caption{\label{fig:flrw}
    Perpendicular magnetic
    field line diffusion coefficients from test
    field line simulations as a function of $B\zero/b_0$ and a theoretically motivated
    approximation (see text) for $s=5/3$. Here
    $D\iso=\left.D_{\perp}\right|_{B\zero=0}$. Note that the
    factor $4 D\iso/\lc \approx 2.4$, obtained here analytically, is not
    expected to depend strongly on $s$.
  }
\end{figure}

We find that for the local perpendicular diffusivity obtained in our
simulations,
the expression
\EQ \label{eq:kperprough}
\kappa_{\perp}(\RL,\eta) = \frac{\kappa\iso(\RL)}{1 + \chi (B\zero/b_0)^2}
\EN
provides a good approximation when $\RL \ll \lc$.
Here $\chi$ is a constant around 2 or 3 that is not very sensitive
to the value of $\RL$ and $\kappa\iso(\RL)$ is given by Equation~(\ref{eq:kappaiso}).
We have arrived at this form by first comparing the magnetic field line
diffusion coefficient $\Db$, as discussed in Section~\ref{section:zeromean},
with $\kappa_{\perp}$, motivated by previous test particle simulations
\citep[e.g.,][]{MichalekOstrowski97,RuffoloEA2008} that have shown
that these two quantities are related. We calculated $\Db$ for various values of
$B\zero/b_0$ using test magnetic line simulations
and compared this with $\kappa_{\perp}$ from our test particle simulations.
For a given $\RL$, our results show approximately $\kappa_{\perp} \propto \Db$
for a large range of $B\zero/b_0$. Then using Equation~(\ref{eq:kappaiso}) we can
write
\EQ \label{eq:kappaDb}
\kappa_{\perp}(\RL,\eta) = \kappa\iso(\RL)\Db(\eta)/D\iso,
\EN
where
$D\iso = \Db|_{B\zero=0}$. As mentioned in Section~\ref{section:zeromean}, $D\iso$ can be
approximated analytically in units of $\lc$ \citep{SonsretteeEA2015}. We have no
general expression for $\Db$, but \citet{SonsretteeEA2015b} have derived analytical
expressions in the limit $B\iso/b_0 \to \infty$ and shown that a simple rational
function which interpolates between $D\iso$ and this limit describes test field line
simulations very well. Following these authors, for large $B\iso/b_0$ we have
$\Db \to D_{\infty} = (\lc/4)(b_0 / B\iso)^2$, then we obtain the expression
\EQ \label{eq:dbapprox}
\Db(\eta) \approx \frac{D\iso}{1 + D\iso/D_{\infty}}
= \frac{D\iso}{1 + (4D\iso/\lc)(B\iso/b_0)^2},
\EN
where we have introduced a rational function that interpolates between
$D\iso$ and $D_{\infty}$. As shown in Figure~\ref{fig:flrw}, this is in very
good agreement with our test field line simulations.
Finally, substituting Equation~(\ref{eq:dbapprox}) into Equation~(\ref{eq:kappaDb}), we
obtain Equation~(\ref{eq:kperprough}) with $\chi = 4 D\iso / \lc$.
If we take $D\iso$ obtained directly from field line simulations
we get $\chi \approx 2.35$, which is very close to the theoretical value $\chi=2.4$ used
in Figure~\ref{fig:flrw}.

We also obtained $D\iso$ from simulation using $s=3/2$; for these,
$\chi \approx2.44$, which only differs from the $s=5/3$ value by a few per cent, which
is about the same size as the error. Moreover, since there are reasonable arguments that $D\iso$
should scale linearly with $\lc$, we expect that $\chi$ should be
essentially a constant, independent of the detailed form of the magnetic spectrum.

In the following subsection we compare Equation~(\ref{eq:kperprough}) and some
improvements directly with our test particle simulation data.

\begin{figure}
  \begin{center}
    \includegraphics[width=0.45\textwidth]{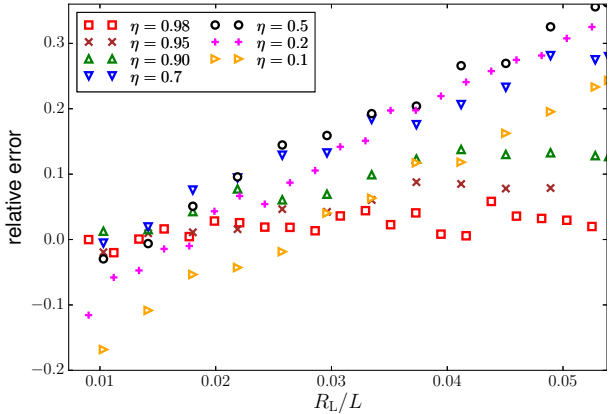}
  \end{center}
  \caption{\label{fig:kperp} The relative error in the fit given by
   Equation~(\ref{eq:kperprough}) (using $\chi=2.35$) to $\kappa_{\perp}$ obtained from test particle
   simulations 
   shown for various values of $\RL$ and $\eta$. The smallest $\RL$ is at the scale
   $2\pi/\kmax$ and the largest at $\lc/2$. Although the errors are quite
   small,
   they vary systematically with $\RL/L$, suggesting that
   there is a missing further dependence on $\RL$.
  }
\end{figure}
\begin{figure}
  \begin{center}
    \includegraphics[width=0.45\textwidth]{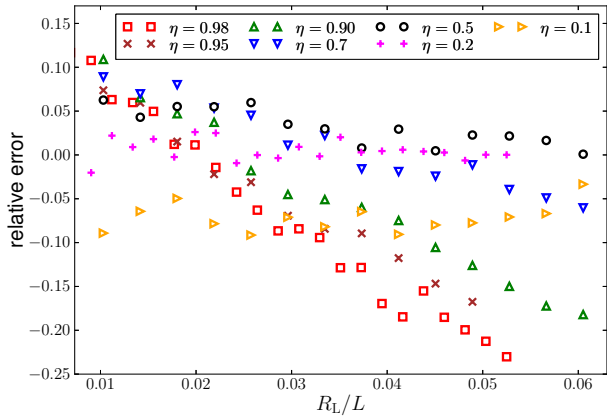}
  \end{center}
  \caption{\label{fig:kperp2} As Figure~\ref{fig:kperp}, but with the approximate
   form of $\kappa_{\perp}$ given by Equation~(\ref{eq:kperppow}).
  }
\end{figure}
\begin{figure}
  \begin{center}
    \includegraphics[width=0.45\textwidth]{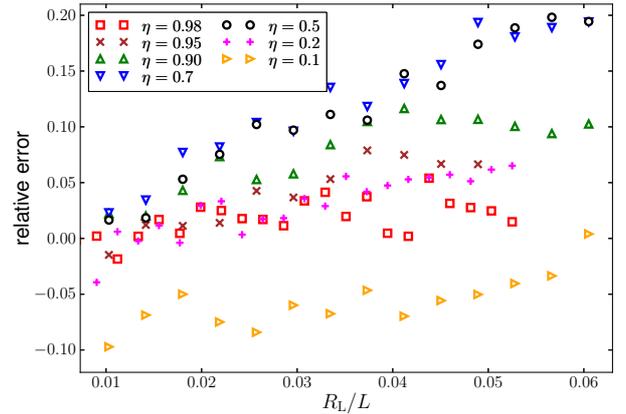}
  \end{center}
  \caption{\label{fig:kperptr} 
   As Figure~\ref{fig:kperp}, but with the
   $\kappa_{\perp}$ given by Equation~(\ref{eq:kperp}).
  }
\end{figure}

\subsection[]{Perpendicular diffusion for $\RL \lesssim \lc/2$}

As shown in Figure~\ref{fig:kperp}, $\chi = 4 D\iso / \lc$ in Equation~(\ref{eq:kperprough})
provides a very good approximation to $\kappa_{\perp}$ obtained from test particle simulations.
However modest in magnitude, the relative error varies systematically with $\RL$, inviting
a more accurate approximation. Figure~\ref{fig:kperp} suggests an additional power-law
factor in $\RL$ with the exponent of 0.54--0.72 for a given $\eta$.
For $\eta=0.2$, we fitted the numerical results in the range $0 \le \RL \le \lc$ to obtain
\EQ \label{eq:kperppow}
\frac{\kappa_{\perp}(\RL,\eta)}{v\zero L} \approx
			\frac{0.19 (\RL/L)^{0.61}}{1 + (4 D\iso/\lc) [\,(1 - \eta)/\eta\,]}\,,
\EN
where the exponent has a standard error of less than one per cent.
For $\eta=0.5$, the numerator becomes $0.20 (\RL/L)^{0.64}$, with a similarly small error.
Furthermore, the numerator exhibits a dependence on $\eta$. The relative difference between the fit at $\eta=0.2$
and the numerical results at different values of $\eta$ is shown in Figure~\ref{fig:kperp2}.
For $\eta\to1$, the difference varies linearly with $\RL/L$, similarly to $\kappa\iso$ in
Equation~\eqref{eq:kappaiso} or the related approximations of Section~\ref{section:zeromean}.
However, variation with a smaller exponent is apparent when $\eta\to0$. The following
heuristic form
with a linear transition between these regimes appears to be a suitable improvement of
Equation~\eqref{eq:kperppow}:
\EQ \label{eq:kperp}
{\frac{\kappa_{\perp}(\RL,\eta)}{v\zero L}
\approx
\frac{\eta \displaystyle\frac{\kappa\iso}{v\zero L}
		+ 0.19\,(1-\eta) \left(\frac{\RL}{L}\right)^{0.61}}{1 +\displaystyle \frac{4 D\iso}{\lc}\,\frac{1 - \eta}{\eta}}\,.}
\EN
The deviations of the numerical results from this form are shown in Figure~\ref{fig:kperptr}.

\section[]{Summary and discussion}

The main outcome of our work is a set of approximate expressions for the diffusion tensor that apply
to cosmic ray propagation in a random magnetic field that has sufficient energy at the
Larmor radius scale to scatter the particles.
The latter condition is not a generic constraint
but rather a result of the limitations of the model
(for example,
our model magnetic field does not include the propagation of magnetohydrodynamic waves,
which should relax this constraint in a physical manner). Our results can be represented as combinations of existing
theoretical ideas and analytical estimates, in forms which may have been difficult to envisage
without detailed computations.

Our main motivation has been to make a first step towards a sub-grid model of cosmic ray propagation in the MHD context,
that captures the dependence of the cosmic ray diffusion on the local physical parameters of the
plasma. We envisage that the approximations to the elements of the diffusion tensor
derived here may be used with the values of
$\eta$ and $L$ obtained from the magnetic field
parameters near the smallest resolved scales of an MHD simulation. It is presumed that the effects
of the magnetic field at scales larger than the numerical resolution would be accounted for by the
solution of equations for cosmic-ray dynamics, e.g., in the advection--diffusion approximation.

The main contribution to the large-scale dynamics of interstellar and intergalactic plasmas
comes from cosmic ray particles in the GeV energy range whose Larmor radius $\RL$ is by far smaller than
any other scale in the MHD approximation. Therefore, a sub-grid model should be based
on a description of the cosmic ray diffusion in the limit $\RL/\delta x\ll1$, where $\delta x$ is
the numerical resolution length. Most studies of cosmic ray diffusion are in the opposite limiting case,
where the Larmor radius is comparable to or larger than the magnetic turbulent scale.
Furthermore, calculations of the diffusion tensor $K_{ij}$ are usually presented in a reference frame
connected with the \textit{global}
mean magnetic field. In our case, such a mean magnetic field is undefined since there are no
reasons to expect any scale separation in the magnetic fields at length scales of a few parsecs.
Therefore, we present our results in terms of the \textit{local} diffusivities $\kappa_{ij}$
defined with respect to the local magnetic field.

We have found the following approximations to the elements of the diffusion cosmic ray tensor
valid when the particles' Larmor radius is much smaller than the scale of magnetic turbulence,
$\RL \ll \lc$, with $\lc$ the correlation length; we defer comparison of the approximations
with previous results till later in this section.

In a purely random magnetic field,
fits to the simulation data in Section~\ref{section:zeromean} suggest that the
isotropic diffusivity can be approximated by
\begin{equation}\label{k_res}
\kappa\iso(\RL)\approx v\zero L\left(a_1 +a_2\frac{\RL}{L}\right)\,,
\end{equation}
where $v\zero$ is the particle speed, $L$ is the outer turbulence scale and
the parameters $a_1$ and $a_2$ depend on the
magnetic power spectrum. The factor $a_2$ changes more weakly with the spectrum than $a_1$, at least
for the spectra explored.
For a power-law spectrum with the exponent $s=5/3$, we have obtained
$a_1\approx3.1\times10^{-3}$ and $a_2\approx0.74$, compared to
$a_1\approx1.9\times10^{-3}$ and $a_2\approx0.76$ for $s=3/2$.
These approximations remain accurate to within about 15 per cent up to $\RL = \lc$.
For relativistic particles ($v\zero = c$) in a field of $5\,\muG$ with $L=100\,\pc$,
these expressions imply a minimum parallel diffusion coefficient of
(2--$3)\times10^{28}\,\cm^2\,\s^{-1}$ (in the limit of a vanishing mean field, $\eta \to 1$).

In a partially ordered magnetic field, with the mean strength $B\zero$ and the root-mean-square
random part $b_0$, the parallel diffusivity of the form \eqref{eq:kpara}, approximates our numerical results for $0 < B_0/b_0 \le 1/3$ to within
about five per cent at $\RL=\lc/4$ (and better at smaller values of $\RL$).
The perpendicular diffusivity is approximated by Equation~\eqref{eq:kperprough}
with $\chi=4D\iso/\lc$,
where $D\iso$ is the isotropic magnetic field line diffusion coefficient (related to the divergence of magnetic lines)
of Equation~\eqref{eq:D} for $\eta=1$, controlled by the integral length scale of the magnetic spectrum,
$D\iso/\lc \approx 0.6$.

Although it is not our primary concern in this paper, we have obtained the following
results for $\RL \lesssim \lc/2$. The perpendicular diffusion coefficient exhibits an
$\RL$-dependence which is not captured by the term containing $\kappa\zero$ in
Equation~\eqref{eq:kperprough}. For $\RL \gtrsim 10^{-3}\lc$, $\kappa_{\perp}$ might be better
approximated by Equation~\eqref{eq:kperppow}. There is a further, weaker dependence on $\eta$, which can
be allowed for using Equation~\eqref{eq:kperp}.
A still better approximation might have the constant $\chi$ in Equation~(\ref{eq:kperprough}) replaced by a function of $\RL$ and/or
$\eta$. Nevertheless, all the given expressions provide a very good
approximation over the range of $\RL$ considered.

To illustrate these results, consider an application to MHD simulations of the ISM adopting,
as typical parameters, $B\zero=2\,\mkG$, $b\zero=5\,\mkG$, $L=100\,\pc$, $s=5/3$ and
$\RL=3.3\times10^{12}\,\cm$. Recent extensive
simulations of the supernova-driven ISM have the numerical resolution of order $\delta x \simeq 1\,\pc$.
The random magnetic field expected at this scale is of the order  of $b|_{\delta x}\simeq b_0(\delta x/L)^{1/3}\simeq1\,\mkG$.
Our intention is to estimate the effective diffusion coefficients at this scale. For this purpose we
can adopt in our diffusion coefficient expressions, $\delta x$ and $b|_{\delta x}$ as the outer scale and fluctuation strength, respectively.
The $\eta$-dependent term in the expression for the parallel local
diffusivity \eqref{eq:kpara} is then of order $10^{-2}$ and is
independent of $\delta x$. This is about three times
larger than the value of $\kappa\iso/(v_0L)|_{L=\delta x}$ estimated in Section~\ref{section:zeromean},
and we find $\kappa_{\parallel} \approx 1.4\times10^{27}\,\cm^2\,\s^{-1}$. From Equation~\eqref{eq:kperprough} we obtain
  $\kappa_{\perp} \approx 3\times10^{25}\,\cm^2\,\s^{-1}$, and $\kappa_{\parallel}/\kappa_{\perp} \approx 50$.
  The random walk of numerically resolved magnetic lines should lead to a more isotropic diffusion tensor
and when applying the usual lengthscale $L$ and fluctuation strength $b_0$ in our expressions we obtain $\kappa_{\parallel}/\kappa_{\perp} \approx 1.5$,
which is in reasonable agreement with the expectation of isotropization at large scales \citep{BerezinskiiEA1990}.
Then the cosmic ray diffusivities
approximated here can be usefully applied to
such MHD simulations
with $b_0$ obtained using the random magnetic field at the smallest numerically resolved scale.

There are other plausible interpretations of the estimated diffusivities. Cosmic ray diffusion
  is controlled at the resonant scales, and a more relevant value of $b_0$ to be used could be that at
the Larmor radius scale. The random magnetic field extrapolated to the Larmor-radius scale,
$b|_{\RL}\simeq b_0(\RL/L)^{1/3}\simeq 4\times10^{-3}\,\mkG$, is very weak indeed.
Since magnetic field at scales exceeding $\delta x$ is fully resolved,
its effect on cosmic ray diffusion does not need to be included into the sub-grid model of
diffusion. Then in place of $B_{0}$, the appropriate effective mean magnetic field could be the
root-mean-square field at the scale $\delta x$, $b_{\delta x}\simeq1\,\mkG$.
Taking these values, the $\RL$-dependent term in the isotropic diffusivity \eqref{k_res} is negligible,
whereas the $\eta$-dependent term in the expression for the parallel local diffusivity
\eqref{eq:kpara} is about 30. Using Equation~\eqref{eq:kperp}, we then obtain
$\kappa_\parallel/\kappa_\perp \simeq10^{7}$. In this scenario, the cosmic ray diffusion tensor
is highly anisotropic at small scales and cosmic rays are partially isotropized when
propagating in the random magnetic field at scales exceeding $\delta x \simeq 1\, \pc$.
Further development and implementation of a sub-grid model for cosmic-ray
diffusion will be discussed elsewhere.

The random magnetic field resolved in MHD simulations of the ISM would enhance
cosmic ray diffusion through the random walk of magnetic lines,
modelled faithfully
by solving simultaneously the MHD and cosmic-ray equations
with the diffusivity tensor dependent on the magnetic field direction.
This process is expected to be more complicated than usually
assumed, and perhaps very different from the predictions of existing
models of cosmic ray propagation in random magnetic fields, because a significant part of the
random magnetic field in the ISM is produced by compression in the shocks driven by
supernova remnants, and by fluctuation dynamo action. Both processes produce
random magnetic fields represented by widely separated, intense structures
against a distributed background; such a random field is strongly intermittent.
The structures associated with the magnetic intermittency are
magnetic sheets and filaments, resulting from shock-wave compression and the fluctuation
dynamo \citep{ZRS90,BBMSS96,WBS07}, respectively. Interstellar
random magnetic fields are expected to have a significant non-Gaussian part, and the
random-phase approximation, as employed here and in several other works, does not apply to such structures.
The magnetic structures can affect the cosmic ray transport significantly by
introducing L\'{e}vy flights of the particles, which would be trapped
for a relatively long time within the structures and move more freely
over larger distances between them. This picture also applies to intergalactic space
where the fluctuation dynamo may be active \citep[e.g.,][]{AR11}.
The fractional volume of magnetic filaments produced in the ISM by the fluctuation
dynamo is estimated to be of order 0.01--0.1 \citep[Sect.~7.4.2 in ][]{S07}.
\citet{FS07} estimate the mean separation between the supernova shocks in the
ISM to be of the order of $10\,\pc$ and argue that polarization observations of the
nearby ISM at low radio frequencies are consistent with this
estimate. It is noteworthy that the localised magnetic structures have scales
by far larger than either the relevant Larmor radii or the resolution of MHD simulations
of the ISM. Therefore, magnetic field models that are free from intermittency, similar to those
used here, are quite appropriate to model cosmic ray diffusion at those small scales,
but not for the global diffusion. The exploration of these effects is
one of the goals of our ongoing simulations of the ISM.

The expressions for the components of the cosmic-ray diffusion tensor obtained here
are broadly consistent with the existing theoretical and computational results. The isotropic diffusion coefficient given by Equation~\eqref{k_res} is the sum of an energy independent (for relativistic particles) and a Bohm-like components.
While previous test particle simulations have shown a Bohm-like scaling \citep[e.g.,][]{CasseEA2001,FatuzzoEA2010,BYL11},
it has been suggested that this scaling only applies down to a certain
value of $\RL$ much smaller than $\lc$. Here we have found, more precisely,
that the scaling continues down to $\RL \approx 2\pi/\kmax$, which seems reasonable. It is not immediately clear why
the previous simulations are inconsistent with this result or how to obtain theoretically the
energy-independent part of the diffusion coefficient. Some previous works \citep[e.g.,][]{GAP08,HMR14}
have fitted isotropic diffusion results by including the theoretically motivated quasilinear $E^{2-s}$ term,
equivalent to the $\RL^{1/3}$ term in our parallel diffusion expression, Equation~\eqref{eq:kpara}. Despite this
difference, Equation~\eqref{k_res} varies less than 15 per cent from the expression of \citet{HMR14} for
$\lc/50 \lesssim \RL < \lc/2$, when $s=5/3$.
The parallel diffusion coefficient given by Equation~\eqref{eq:kpara} appears to be
consistent with a quasilinear scaling at low $\eta$, while the simple addition of the isotropic diffusion coefficient
compensates very well for the breakdown of this scaling at $\eta \to 1$. One should note that the quasilinear result
of Equation~\eqref{eq:kparaQL} is based on an MHD wave propagation model in the limit of
zero-frequency waves \citep[][]{BerezinskiiEA1990,AloisioBerezinsky2004}, whereas a calculation
more consistent with our underlying magnetic field model, i.e., magnetostatic turbulence, usually yields an infinite diffusion coefficient \citep{FiskEA1974}.
[This is a problem with the original quasilinear theory, relating to scattering at pitch angles of $\theta=\pi/2$. The problem might be overcome in several ways;
see e.g.\ \cite{CesarskyKulsrud1973}, \cite{TautzEA2008}]. One explanation for this apparent good agreement with a slightly inconsistent theory might be that the effect of
propagating MHD waves is not very important when considering asymptotic diffusion coefficients \citep{FatuzzoEA2010}.
And finally, the local perpendicular diffusivity $\kappa_\perp$ scales
almost perfectly with the magnetic field line diffusion coefficient.

Regarding test particle simulations, we have shown that discrete and continuous
  magnetic field constructions can yield essentially identical results over a broad range of
  particle energies. For magnetic fields on a discrete mesh, our results suggest that caution is required
when $\RL$ is less than a few mesh points or is a significant fraction of $L$.

\section*{Acknowledgments}
AS and LFSR acknowledge useful discussions with the members of the International Team~323
of the International Space Science Institute in Bern.
Useful comments of Diego Harari are gratefully acknowledged.
APS would like to thank David Ruffolo and Wirin Sonsrettee for useful discussions.
APS is grateful to the School of Mathematics and Statistics in Newcastle for hospitality.
This work was supported by
the Leverhulme Trust (Grant RPG-2014-427) and the STFC (Grant ST/L005549/1).

\footnotesize{
  \bibliographystyle{mnras}
  \bibliography{crdiffusion}

\begin{thebibliography}{}
\makeatletter
\relax
\def\mn@urlcharsother{\let\do\@makeother \do\$\do\&\do\#\do\^\do\_\do\%\do\~}
\def\mn@doi{\begingroup\mn@urlcharsother \@ifnextchar [ {\mn@doi@}
  {\mn@doi@[]}}
\def\mn@doi@[#1]#2{\def\@tempa{#1}\ifx\@tempa\@empty \href
  {http://dx.doi.org/#2} {doi:#2}\else \href {http://dx.doi.org/#2} {#1}\fi
  \endgroup}
\def\mn@eprint#1#2{\mn@eprint@#1:#2::\@nil}
\def\mn@eprint@arXiv#1{\href {http://arxiv.org/abs/#1} {{\tt arXiv:#1}}}
\def\mn@eprint@dblp#1{\href {http://dblp.uni-trier.de/rec/bibtex/#1.xml}
  {dblp:#1}}
\def\mn@eprint@#1:#2:#3:#4\@nil{\def\@tempa {#1}\def\@tempb {#2}\def\@tempc
  {#3}\ifx \@tempc \@empty \let \@tempc \@tempb \let \@tempb \@tempa \fi \ifx
  \@tempb \@empty \def\@tempb {arXiv}\fi \@ifundefined
  {mn@eprint@\@tempb}{\@tempb:\@tempc}{\expandafter \expandafter \csname
  mn@eprint@\@tempb\endcsname \expandafter{\@tempc}}}

\bibitem[\protect\citeauthoryear{{Akahori} \& {Ryu}}{{Akahori} \&
  {Ryu}}{2011}]{AR11}
{Akahori} T.,  {Ryu} D.,  2011, \mn@doi [\apj] {10.1088/0004-637X/738/2/134},
  \href {http://adsabs.harvard.edu/abs/2011ApJ...738..134A} {738, 134}

\bibitem[\protect\citeauthoryear{{Aloisio} \& {Berezinsky}}{{Aloisio} \&
  {Berezinsky}}{2004}]{AloisioBerezinsky2004}
{Aloisio} R.,  {Berezinsky} V.,  2004, \mn@doi [\apj] {10.1086/421869}, \href
  {http://adsabs.harvard.edu/abs/2004ApJ...612..900A} {612, 900}

\bibitem[\protect\citeauthoryear{{Armstrong}, {Rickett}  \&
  {Spangler}}{{Armstrong} et~al.}{1995}]{ARS95}
{Armstrong} J.~W.,  {Rickett} B.~J.,   {Spangler} S.~R.,  1995, \mn@doi [\apj]
  {10.1086/175515}, \href {http://adsabs.harvard.edu/abs/1995ApJ...443..209A}
  {443, 209}

\bibitem[\protect\citeauthoryear{{Bakunin}}{{Bakunin}}{2008}]{B08}
{Bakunin} O.~G.,  2008, Turbulence and Diffusion: Scaling Versus Equations.
Springer, Berlin

\bibitem[\protect\citeauthoryear{{Batchelor}}{{Batchelor}}{1953}]{Batchelor1953}
{Batchelor} G.~K.,  1953, {The Theory of Homogeneous Turbulence}.
Cambridge University Press

\bibitem[\protect\citeauthoryear{{Beck}, {Brandenburg}, {Moss}, {Shukurov}  \&
  {Sokoloff}}{{Beck} et~al.}{1996}]{BBMSS96}
{Beck} R.,  {Brandenburg} A.,  {Moss} D.,  {Shukurov} A.,   {Sokoloff} D.,
  1996, \mn@doi [\araa] {10.1146/annurev.astro.34.1.155}, \href
  {http://adsabs.harvard.edu/abs/1996ARA%26A..34..155B} {34, 155}

\bibitem[\protect\citeauthoryear{{Beck}, {Shukurov}, {Sokoloff}  \&
  {Wielebinski}}{{Beck} et~al.}{2003}]{BSSW03}
{Beck} R.,  {Shukurov} A.,  {Sokoloff} D.,   {Wielebinski} R.,  2003, \mn@doi
  [\aap] {10.1051/0004-6361:20031101}, \href
  {http://adsabs.harvard.edu/abs/2003A%26A...411...99B} {411, 99}

\bibitem[\protect\citeauthoryear{{Beresnyak}, {Yan}  \& {Lazarian}}{{Beresnyak}
  et~al.}{2011}]{BYL11}
{Beresnyak} A.,  {Yan} H.,   {Lazarian} A.,  2011, \mn@doi [\apj]
  {10.1088/0004-637X/728/1/60}, \href
  {http://adsabs.harvard.edu/abs/2011ApJ...728...60B} {728, 60}

\bibitem[\protect\citeauthoryear{{Berezinskii}, {Bulanov}, {Dogiel}  \&
  {Ptuskin}}{{Berezinskii} et~al.}{1990}]{BerezinskiiEA1990}
{Berezinskii} V.~S.,  {Bulanov} S.~V.,  {Dogiel} V.~A.,   {Ptuskin} V.~S.,
  1990, {Astrophysics of Cosmic Rays}.
North-Holland, Amsterdam

\bibitem[\protect\citeauthoryear{{Boris}}{{Boris}}{1970}]{Boris70}
{Boris} J.~P.,  1970, in Proceedings of the Fourth Conference on Numerical
  Simulation of Plasmas. {Naval Res. Lab., Washington, D.C}, pp 3--76

\bibitem[\protect\citeauthoryear{{Candia} \& {Roulet}}{{Candia} \&
  {Roulet}}{2004}]{CandiaRoulet2004}
{Candia} J.,  {Roulet} E.,  2004, \mn@doi [\jcap]
  {10.1088/1475-7516/2004/10/007}, \href
  {http://adsabs.harvard.edu/abs/2004JCAP...10..007C} {10, 7}

\bibitem[\protect\citeauthoryear{{Cash} \& {Karp}}{{Cash} \&
  {Karp}}{1990}]{CashKarp90}
{Cash} J.~R.,  {Karp} A.~H.,  1990, \mn@doi [ACM Trans. Math. Softw.]
  {10.1145/79505.79507}, 16, 201

\bibitem[\protect\citeauthoryear{Casse, Lemoine  \& Pelletier}{Casse
  et~al.}{2001}]{CasseEA2001}
Casse F.,  Lemoine M.,   Pelletier G.,  2001, \mn@doi [Phys.\ Rev.\ D]
  {10.1103/PhysRevD.65.023002}, 65, 023002

\bibitem[\protect\citeauthoryear{{Cesarsky} \& {Kulsrud}}{{Cesarsky} \&
  {Kulsrud}}{1973}]{CesarskyKulsrud1973}
{Cesarsky} C.~J.,  {Kulsrud} R.~M.,  1973, \mn@doi [\apj] {10.1086/152405},
  \href {http://adsabs.harvard.edu/abs/1973ApJ...185..153C} {185, 153}

\bibitem[\protect\citeauthoryear{{De Marco}, {Blasi}  \& {Stanev}}{{De Marco}
  et~al.}{2007}]{DeMarcoEA2007}
{De Marco} D.,  {Blasi} P.,   {Stanev} T.,  2007, \mn@doi [\jcap]
  {10.1088/1475-7516/2007/06/027}, \href
  {http://adsabs.harvard.edu/abs/2007JCAP...06..027D} {6, 27}

\bibitem[\protect\citeauthoryear{{Evoli}, {Gaggero}, {Grasso}  \&
  {Maccione}}{{Evoli} et~al.}{2008}]{EGGM08}
{Evoli} C.,  {Gaggero} D.,  {Grasso} D.,   {Maccione} L.,  2008, \mn@doi
  [\jcap] {10.1088/1475-7516/2008/10/018}, \href
  {http://adsabs.harvard.edu/abs/2008JCAP...10..018E} {10, 18}

\bibitem[\protect\citeauthoryear{{Farge} \& {Schneider}}{{Farge} \&
  {Schneider}}{2015}]{FS15}
{Farge} M.,  {Schneider} K.,  2015, \mn@doi [J.\ Plasma Phys.]
  {10.1017/S0022377815001075}, \href
  {http://adsabs.harvard.edu/abs/2015arXiv150805650F} {81, 43}

\bibitem[\protect\citeauthoryear{{Fatuzzo}, {Melia}, {Todd}  \&
  {Adams}}{{Fatuzzo} et~al.}{2010}]{FatuzzoEA2010}
{Fatuzzo} M.,  {Melia} F.,  {Todd} E.,   {Adams} F.~C.,  2010, \mn@doi [\apj]
  {10.1088/0004-637X/725/1/515}, \href
  {http://adsabs.harvard.edu/abs/2010ApJ...725..515F} {725, 515}

\bibitem[\protect\citeauthoryear{{Fisk}, {Goldstein}, {Klimas}  \&
  {Sandri}}{{Fisk} et~al.}{1974}]{FiskEA1974}
{Fisk} L.~A.,  {Goldstein} M.~L.,  {Klimas} A.~J.,   {Sandri} G.,  1974,
  \mn@doi [\apj] {10.1086/152893}, \href
  {http://adsabs.harvard.edu/abs/1974ApJ...190..417F} {190, 417}

\bibitem[\protect\citeauthoryear{{Fletcher} \& {Shukurov}}{{Fletcher} \&
  {Shukurov}}{2007}]{FS07}
{Fletcher} A.,  {Shukurov} A.,  2007, in {Miville-Desch{\^e}nes} M.-A.,
  {Boulanger} F.,  eds,  EAS Publ.\ Ser. Vol. 23, {Sky Polarisation at
  Far-Infrared to Radio Wavelengths: The Galactic Screen before the Cosmic
  Microwave Background}. pp 109--128 (\mn@eprint {} {astro-ph/0602536})

\bibitem[\protect\citeauthoryear{{Fung}, {Hunt}, {Malik}  \& {Perkins}}{{Fung}
  et~al.}{1992}]{FungEA1992}
{Fung} J.~C.~H.,  {Hunt} J.~C.~R.,  {Malik} N.~A.,   {Perkins} R.~J.,  1992,
  \mn@doi [\jfm] {10.1017/S0022112092001423}, 236, 281

\bibitem[\protect\citeauthoryear{{Gaggero}, {Maccione}, {Di Bernardo}, {Evoli}
  \& {Grasso}}{{Gaggero} et~al.}{2013}]{GMDEG13}
{Gaggero} D.,  {Maccione} L.,  {Di Bernardo} G.,  {Evoli} C.,   {Grasso} D.,
  2013, \mn@doi [\prl] {10.1103/PhysRevLett.111.021102}, \href
  {http://adsabs.harvard.edu/abs/2013PhRvL.111b1102G} {111, 021102}

\bibitem[\protect\citeauthoryear{{Gent}, {Shukurov}, {Sarson}, {Fletcher}  \&
  {Mantere}}{{Gent} et~al.}{2013}]{GSSFL13b}
{Gent} F.~A.,  {Shukurov} A.,  {Sarson} G.~R.,  {Fletcher} A.,   {Mantere}
  M.~J.,  2013, \mn@doi [\mnras] {10.1093/mnrasl/sls042}, \href
  {http://adsabs.harvard.edu/abs/2013MNRAS.430L..40G} {430, L40}

\bibitem[\protect\citeauthoryear{{Giacalone} \& {Jokipii}}{{Giacalone} \&
  {Jokipii}}{1994}]{GiacaloneJokipii1994}
{Giacalone} J.,  {Jokipii} J.~R.,  1994, \mn@doi [\apjl] {10.1086/187457},
  \href {http://adsabs.harvard.edu/abs/1994ApJ...430L.137G} {430, L137}

\bibitem[\protect\citeauthoryear{{Giacalone} \& {Jokipii}}{{Giacalone} \&
  {Jokipii}}{1999}]{GiacaloneJokipii1999}
{Giacalone} J.,  {Jokipii} J.~R.,  1999, \mn@doi [\apj] {10.1086/307452}, \href
  {http://adsabs.harvard.edu/abs/1999ApJ...520..204G} {520, 204}

\bibitem[\protect\citeauthoryear{{Globus}, {Allard}  \& {Parizot}}{{Globus}
  et~al.}{2008}]{GAP08}
{Globus} N.,  {Allard} D.,   {Parizot} E.,  2008, \mn@doi [\aap]
  {10.1051/0004-6361:20078653}, \href
  {http://adsabs.harvard.edu/abs/2008A%26A...479...97G} {479, 97}

\bibitem[\protect\citeauthoryear{{Goldstein}, {Roberts}  \&
  {Matthaeus}}{{Goldstein} et~al.}{1995}]{GRM95}
{Goldstein} M.~L.,  {Roberts} D.~A.,   {Matthaeus} W.~H.,  1995, \mn@doi
  [\araa] {10.1146/annurev.aa.33.090195.001435}, \href
  {http://adsabs.harvard.edu/abs/1995ARA%26A..33..283G} {33, 283}

\bibitem[\protect\citeauthoryear{{Green}}{{Green}}{1951}]{Green51}
{Green} M.~S.,  1951, \mn@doi [\jcp] {10.1063/1.1748449}, \href
  {http://adsabs.harvard.edu/abs/1951JChPh..19.1036G} {19, 1036}

\bibitem[\protect\citeauthoryear{{Hanasz}, {W{\'o}lta{\'n}ski}  \&
  {Kowalik}}{{Hanasz} et~al.}{2009}]{HWK09}
{Hanasz} M.,  {W{\'o}lta{\'n}ski} D.,   {Kowalik} K.,  2009, \mn@doi [\apjl]
  {10.1088/0004-637X/706/1/L155}, \href
  {http://adsabs.harvard.edu/abs/2009ApJ...706L.155H} {706, L155}

\bibitem[\protect\citeauthoryear{{Harari}, {Mollerach}  \& {Roulet}}{{Harari}
  et~al.}{2014}]{HMR14}
{Harari} D.,  {Mollerach} S.,   {Roulet} E.,  2014, \mn@doi [\prd]
  {10.1103/PhysRevD.89.123001}, \href
  {http://adsabs.harvard.edu/abs/2014PhRvD..89l3001H} {89, 123001}

\bibitem[\protect\citeauthoryear{{Haverkorn}, {Brown}, {Gaensler}  \&
  {McClure-Griffiths}}{{Haverkorn} et~al.}{2008}]{HBGM08}
{Haverkorn} M.,  {Brown} J.~C.,  {Gaensler} B.~M.,   {McClure-Griffiths} N.~M.,
   2008, \mn@doi [\apj] {10.1086/587165}, \href
  {http://adsabs.harvard.edu/abs/2008ApJ...680..362H} {680, 362}

\bibitem[\protect\citeauthoryear{{Howes}}{{Howes}}{2015}]{H15}
{Howes} G.~G.,  2015, \mn@doi [{Phil. Trans. R. Soc. A}]
  {10.1098/rsta.2014.0145}, \href
  {http://adsabs.harvard.edu/abs/2015RSPTA.37340145H} {373, 40145}

\bibitem[\protect\citeauthoryear{{Iacobelli} et~al.,}{{Iacobelli}
  et~al.}{2013}]{Ietal13}
{Iacobelli} M.,  et~al., 2013, \mn@doi [\aap] {10.1051/0004-6361/201322013},
  \href {http://adsabs.harvard.edu/abs/2013A%26A...558A..72I} {558, A72}

\bibitem[\protect\citeauthoryear{{Jokipii}}{{Jokipii}}{1966}]{Jokipii1966}
{Jokipii} J.~R.,  1966, \mn@doi [\apj] {10.1086/148912}, \href
  {http://adsabs.harvard.edu/abs/1966ApJ...146..480J} {146, 480}

\bibitem[\protect\citeauthoryear{{Jokipii} \& {Parker}}{{Jokipii} \&
  {Parker}}{1968}]{JokipiiParker1968}
{Jokipii} J.~R.,  {Parker} E.~N.,  1968, \mn@doi [Physical Review Letters]
  {10.1103/PhysRevLett.21.44}, \href
  {http://adsabs.harvard.edu/abs/1968PhRvL..21...44J} {21, 44}

\bibitem[\protect\citeauthoryear{{K{\'o}ta} \& {Jokipii}}{{K{\'o}ta} \&
  {Jokipii}}{2000}]{KotaJokipii2000}
{K{\'o}ta} J.,  {Jokipii} J.~R.,  2000, \mn@doi [\apj] {10.1086/308492}, \href
  {http://adsabs.harvard.edu/abs/2000ApJ...531.1067K} {531, 1067}

\bibitem[\protect\citeauthoryear{{Kubo}}{{Kubo}}{1957}]{Kubo57}
{Kubo} R.,  1957, \mn@doi [J.\ Phys.\ Soc.\ Japan] {10.1143/JPSJ.12.570}, \href
  {http://adsabs.harvard.edu/abs/1957JPSJ...12..570K} {12, 570}

\bibitem[\protect\citeauthoryear{{Kulsrud}}{{Kulsrud}}{2004}]{Ku04}
{Kulsrud} R.~M.,  2004, {Plasma Physics for Astrophysics}.
Princeton. Univ.\ Press, Princeton

\bibitem[\protect\citeauthoryear{{Maccione}}{{Maccione}}{2013}]{Ma13}
{Maccione} L.,  2013, \mn@doi [\prl] {10.1103/PhysRevLett.110.081101}, \href
  {http://adsabs.harvard.edu/abs/2013PhRvL.110h1101M} {110, 081101}

\bibitem[\protect\citeauthoryear{{Mace}, {Dalena}  \& {Matthaeus}}{{Mace}
  et~al.}{2012}]{Mace12}
{Mace} R.~L.,  {Dalena} S.,   {Matthaeus} W.~H.,  2012, \mn@doi [Physics of
  Plasmas] {10.1063/1.3693379}, \href
  {http://adsabs.harvard.edu/abs/2012PhPl...19c2309M} {19, 032309}

\bibitem[\protect\citeauthoryear{{Michalek} \& {Ostrowski}}{{Michalek} \&
  {Ostrowski}}{1997}]{MichalekOstrowski97}
{Michalek} G.,  {Ostrowski} M.,  1997, \aap, \href
  {http://adsabs.harvard.edu/abs/1997A%26A...326..793M} {326, 793}

\bibitem[\protect\citeauthoryear{Miniati}{Miniati}{2001}]{Miniati01}
Miniati F.,  2001, \mn@doi [Computer Phys.\ Comm.]
  {10.1016/S0010-4655(01)00293-4}, 141, 17

\bibitem[\protect\citeauthoryear{Miniati}{Miniati}{2007}]{Miniati07}
Miniati F.,  2007, \mn@doi [J.\ Computational Phys.]
  {10.1016/j.jcp.2007.08.013}, 227, 776

\bibitem[\protect\citeauthoryear{{Minter} \& {Spangler}}{{Minter} \&
  {Spangler}}{1996}]{MS96}
{Minter} A.~H.,  {Spangler} S.~R.,  1996, \mn@doi [\apj] {10.1086/176803},
  \href {http://adsabs.harvard.edu/abs/1996ApJ...458..194M} {458, 194}

\bibitem[\protect\citeauthoryear{{Monin} \& {Yaglom}}{{Monin} \&
  {Yaglom}}{1975}]{MoninYaglom1975}
{Monin} A.~S.,  {Yaglom} A.~M.,  1975, {Statistical Fluid Mechanics, Vol.\ 2}.
MIT press, Cambridge, MA

\bibitem[\protect\citeauthoryear{{Parizot}}{{Parizot}}{2004}]{Parizot2004}
{Parizot} E.,  2004, \mn@doi [Nuclear Phys.\ B, Proceedings Suppl.]
  {10.1016/j.nuclphysbps.2004.10.034}, \href
  {http://adsabs.harvard.edu/abs/2004NuPhS.136..169P} {136, 169}

\bibitem[\protect\citeauthoryear{{Plotnikov}, {Pelletier}  \&
  {Lemoine}}{{Plotnikov} et~al.}{2011}]{PlotnikovEA2011}
{Plotnikov} I.,  {Pelletier} G.,   {Lemoine} M.,  2011, \mn@doi [\aap]
  {10.1051/0004-6361/201117182}, \href
  {http://adsabs.harvard.edu/abs/2011A%26A...532A..68P} {532, A68}

\bibitem[\protect\citeauthoryear{{Ruffolo}, {Chuychai}, {Wongpan}, {Minnie},
  {Bieber}  \& {Matthaeus}}{{Ruffolo} et~al.}{2008}]{RuffoloEA2008}
{Ruffolo} D.,  {Chuychai} P.,  {Wongpan} P.,  {Minnie} J.,  {Bieber} J.~W.,
  {Matthaeus} W.~H.,  2008, \mn@doi [\apj] {10.1086/591493}, \href
  {http://adsabs.harvard.edu/abs/2008ApJ...686.1231R} {686, 1231}

\bibitem[\protect\citeauthoryear{{Ruzmaikin}, {Shukurov}  \&
  {Sokoloff}}{{Ruzmaikin} et~al.}{1988}]{RSS88}
{Ruzmaikin} A.~A.,  {Shukurov} A.~M.,   {Sokoloff} D.~D.,  1988, Magnetic
  Fields of Galaxies.
Kluwer, Dordrecht

\bibitem[\protect\citeauthoryear{{Schlickeiser}}{{Schlickeiser}}{2002}]{Sch02}
{Schlickeiser} R.,  2002, {Cosmic Ray Astrophysics}.
Springer, Berlin

\bibitem[\protect\citeauthoryear{{Shalchi}}{{Shalchi}}{2009}]{Sh09}
{Shalchi} A.,  2009, {Nonlinear Cosmic Ray Diffusion Theories}.
~ Vol. 362, Springer, Berlin, \mn@doi{10.1007/978-3-642-00309-7}

\bibitem[\protect\citeauthoryear{{Shalchi}}{{Shalchi}}{2011}]{Sh11}
{Shalchi} A.,  2011, \mn@doi [Phys.\ Rev.\ E] {10.1103/PhysRevE.83.046402},
  \href {http://adsabs.harvard.edu/abs/2011PhRvE..83d6402S} {83, 046402}

\bibitem[\protect\citeauthoryear{Shalchi \& Dosch}{Shalchi \&
  Dosch}{2009}]{SD09}
Shalchi A.,  Dosch A.,  2009, \mn@doi [Phys.\ Rev.\ D]
  {10.1103/PhysRevD.79.083001}, 79, 083001

\bibitem[\protect\citeauthoryear{{Shukurov}}{{Shukurov}}{2007}]{S07}
{Shukurov} A.,  2007, in {Dormy} E.,  {Soward} A.~M.,  eds, Mathematical
  Aspects of Natural Dynamos. Chapman \& Hall/CRC, London, pp 313--359

\bibitem[\protect\citeauthoryear{{Siejkowski}, {Otmianowska-Mazur}, {Soida},
  {Bomans}  \& {Hanasz}}{{Siejkowski} et~al.}{2014}]{SOSBH14}
{Siejkowski} H.,  {Otmianowska-Mazur} K.,  {Soida} M.,  {Bomans} D.~J.,
  {Hanasz} M.,  2014, \mn@doi [\aap] {10.1051/0004-6361/201220367}, \href
  {http://adsabs.harvard.edu/abs/2014A%26A...562A.136S} {562, A136}

\bibitem[\protect\citeauthoryear{{Snodin}, {Brandenburg}, {Mee}  \&
  {Shukurov}}{{Snodin} et~al.}{2006}]{SBMS06}
{Snodin} A.~P.,  {Brandenburg} A.,  {Mee} A.~J.,   {Shukurov} A.,  2006,
  \mn@doi [\mnras] {10.1111/j.1365-2966.2006.11034.x}, \href
  {http://adsabs.harvard.edu/abs/2006MNRAS.373..643S} {373, 643}

\bibitem[\protect\citeauthoryear{{Snodin}, {Ruffolo}  \& {Matthaeus}}{{Snodin}
  et~al.}{2013}]{SRM13}
{Snodin} A.~P.,  {Ruffolo} D.,   {Matthaeus} W.~H.,  2013, \mn@doi [\apj]
  {10.1088/0004-637X/762/1/66}, \href
  {http://adsabs.harvard.edu/abs/2013ApJ...762...66S} {762, 66}

\bibitem[\protect\citeauthoryear{Sonsrettee, Subedi, Ruffolo, Matthaeus,
  Snodin, Wongpan  \& Chuychai}{Sonsrettee et~al.}{2015a}]{SonsretteeEA2015}
Sonsrettee W.,  Subedi P.,  Ruffolo D.,  Matthaeus W.~H.,  Snodin A.~P.,
  Wongpan P.,   Chuychai P.,  2015a, \mn@doi [\apj]
  {10.1088/0004-637X/798/1/59}, 798, 59

\bibitem[\protect\citeauthoryear{Sonsrettee et~al.,}{Sonsrettee
  et~al.}{2015b}]{SonsretteeEA2015b}
Sonsrettee W.,  et~al., 2015b, {submitted to \apjs\ (2015)}

\bibitem[\protect\citeauthoryear{{Stepanov}, {Shukurov}, {Fletcher}, {Beck},
  {La Porta}  \& {Tabatabaei}}{{Stepanov} et~al.}{2014}]{SSFTLT14}
{Stepanov} R.,  {Shukurov} A.,  {Fletcher} A.,  {Beck} R.,  {La Porta} L.,
  {Tabatabaei} F.,  2014, \mn@doi [\mnras] {10.1093/mnras/stt2044}, \href
  {http://adsabs.harvard.edu/abs/2014MNRAS.437.2201S} {437, 2201}

\bibitem[\protect\citeauthoryear{{Strong} \& {Moskalenko}}{{Strong} \&
  {Moskalenko}}{2001}]{SM01}
{Strong} A.~W.,  {Moskalenko} I.~V.,  2001, \mn@doi [Adv.\ Space Res.]
  {10.1016/S0273-1177(01)00112-0}, \href
  {http://adsabs.harvard.edu/abs/2001AdSpR..27..717S} {27, 717}

\bibitem[\protect\citeauthoryear{{Strong}, {Moskalenko}  \& {Ptuskin}}{{Strong}
  et~al.}{2007}]{SMP07}
{Strong} A.~W.,  {Moskalenko} I.~V.,   {Ptuskin} V.~S.,  2007, \mn@doi [Ann.\
  Rev.\ Nucl.\ Particle Sci.] {10.1146/annurev.nucl.57.090506.123011}, \href
  {http://adsabs.harvard.edu/abs/2007ARNPS..57..285S} {57, 285}

\bibitem[\protect\citeauthoryear{{Tautz}, {Shalchi}  \& {Schlickeiser}}{{Tautz}
  et~al.}{2008}]{TautzEA2008}
{Tautz} R.~C.,  {Shalchi} A.,   {Schlickeiser} R.,  2008, \mn@doi [\apjl]
  {10.1086/592498}, \href {http://adsabs.harvard.edu/abs/2008ApJ...685L.165T}
  {685, L165}

\bibitem[\protect\citeauthoryear{{Taylor}}{{Taylor}}{1922}]{Taylor22}
{Taylor} G.~I.,  1922, \mn@doi [Proc.\ London Math.\ Soc.]
  {10.1112/plms/s2-20.1.196}, s2-20, 196

\bibitem[\protect\citeauthoryear{{Urch}}{{Urch}}{1977}]{Urch1977}
{Urch} I.~H.,  1977, \mn@doi [\apss] {10.1007/BF00644386}, \href
  {http://adsabs.harvard.edu/abs/1977Ap%26SS..46..389U} {46, 389}

\bibitem[\protect\citeauthoryear{{Vladimirov} et~al.,}{{Vladimirov}
  et~al.}{2011}]{VDJMMNOPS11}
{Vladimirov} A.~E.,  et~al., 2011, \mn@doi [Computer Phys.\ Comm.]
  {10.1016/j.cpc.2011.01.017}, \href
  {http://adsabs.harvard.edu/abs/2011CoPhC.182.1156V} {182, 1156}

\bibitem[\protect\citeauthoryear{{Wilkin}, {Barenghi}  \& {Shukurov}}{{Wilkin}
  et~al.}{2007}]{WBS07}
{Wilkin} S.~L.,  {Barenghi} C.~F.,   {Shukurov} A.,  2007, \mn@doi [Phys.\
  Rev.\ Lett.] {10.1103/PhysRevLett.99.134501}, \href
  {http://adsabs.harvard.edu/abs/2007PhRvL..99m4501W} {99, 134501}

\bibitem[\protect\citeauthoryear{{Zeldovich}, {Ruzmaikin}  \&
  {Sokoloff}}{{Zeldovich} et~al.}{1990}]{ZRS90}
{Zeldovich} Y.~B.,  {Ruzmaikin} A.~A.,   {Sokoloff} D.~D.,  1990, {The Almighty
  Chance}.
World Scientific Publication, Singapore

\makeatother
\end{thebibliography}
}

\appendix

\section[]{Comparison of magnetic field models and resolution tests}\label{results:comparison}\label{AppA}
\begin{figure}
  \begin{center}
    \includegraphics[width=0.4\textwidth]{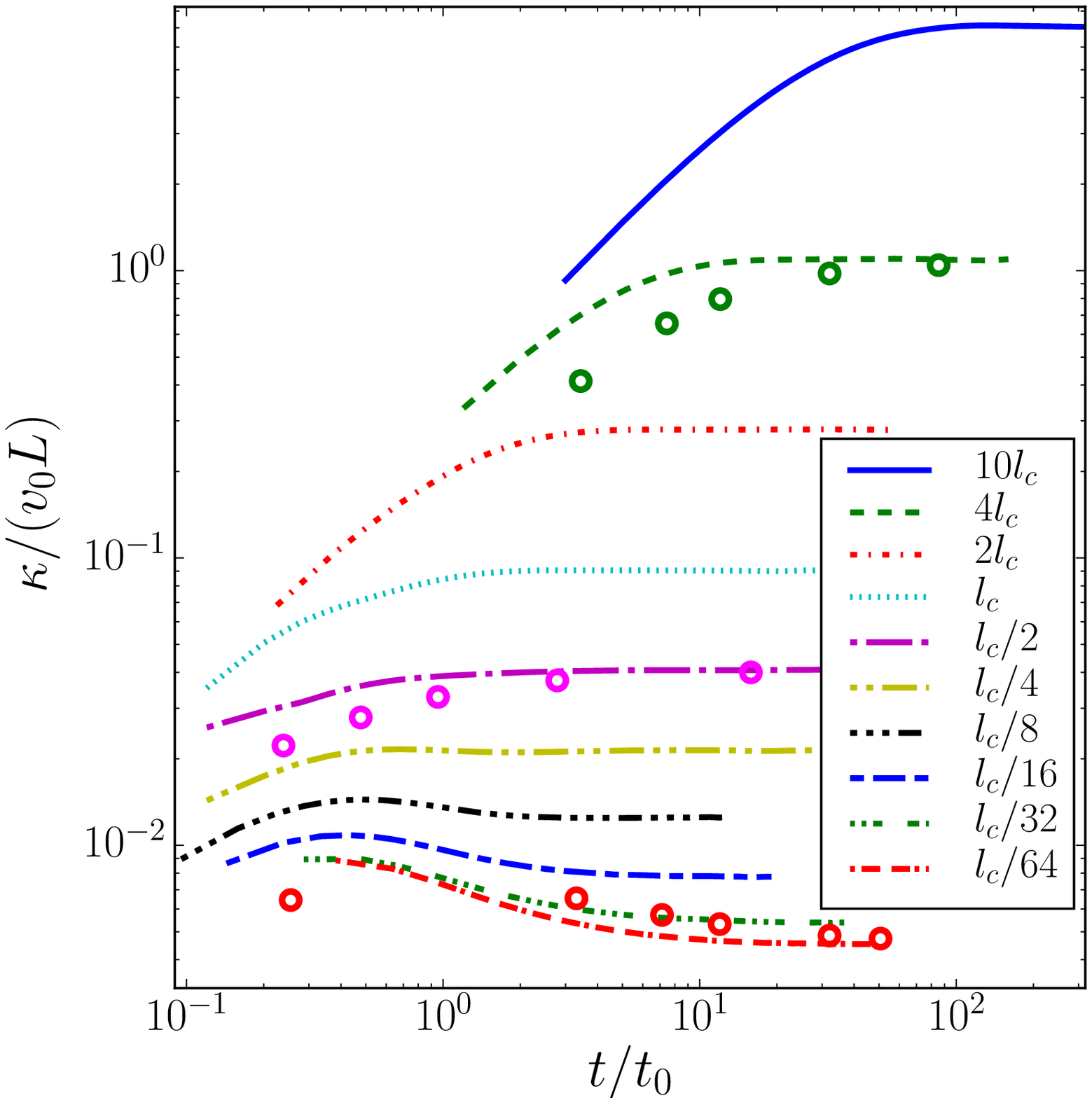}
  \end{center}
  \caption{\label{evo} Time evolution of the (isotropic) running diffusion
    coefficient for vanishing mean field, $B_0=0$, for various values of $\RL$
   as specified in the legend. The continuous magnetic field model
 has been used with $N=512$ and $\kmax/k\zero = 256$. In the case of
 $\RL=4\lc,\lc/2$ and $\lc/64$, open circles show the evaluation of the usual
 asymptotic diffusion coefficient, i.e.\ Equation~(\ref{Partial}), after time $t/t_0$.
 We see that this is consistent with the running diffusion coefficient if
 evaluated at sufficiently large time. However, for practical evaluation of the
 diffusion coefficients, clearly the running diffusion coefficient is preferable,
 since it converges at much earlier time.
 }
\end{figure}

\begin{figure}
  \begin{center}
    \includegraphics[width=0.45\textwidth]{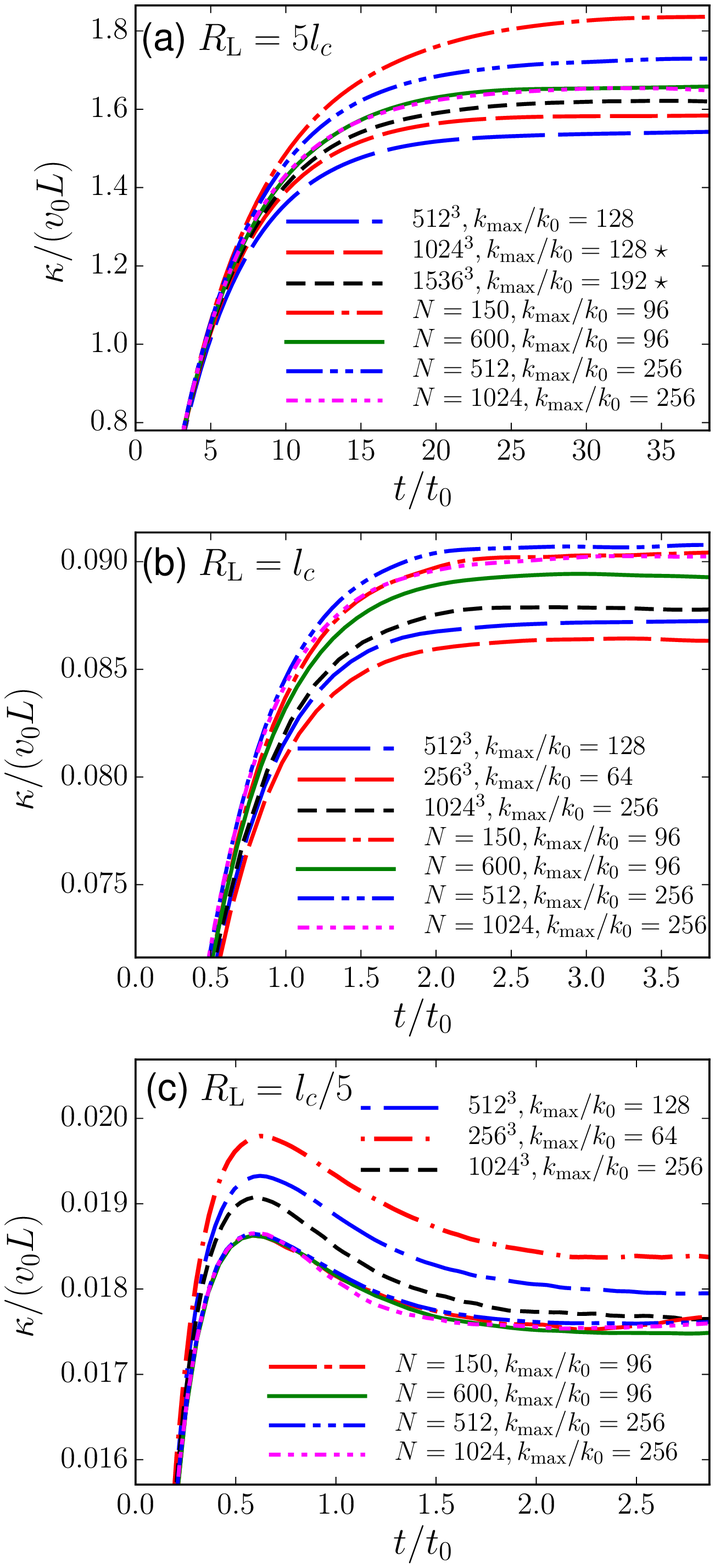}
  \end{center}
  \caption{\label{comparemethods}
  A comparison of the running diffusion coefficient obtained for the continuous
  and discrete magnetic field models under various numerical resolutions in
  $k$-space, specified in the legends, for $s=5/3$ and a purely random magnetic
  field, $B\zero=0$, at various values of $\RL$: (a) $\RL =5 \lc$; (b) $\RL=\lc$ and (c) $\RL = \lc / 5$.
    For the discrete model \eqref{bkmesh}, $k\zero$ is taken to be equal to
    two mesh separations in wave number space (see page~\pageref{subtleties}),
    apart from the two cases denoted with a star
    in Panel (a), for which $k\zero$
    is equal to four mesh separations.
    }
\end{figure}

First, for reference, Figure~\ref{evo} compares the time evolution of $\kappa\iso$
given by Equation~(\ref{Running}) at various values of $\RL$.
For $\lc/8 \le \RL \le 2\lc$, the subdiffusive regime is rather short (about ten time units)
and the diffusivity quickly reaches an asymptotic value.
For $\RL \gg \lc$, both the diffusion coefficients and the time required to reach the
diffusive propagation can be very large. With open circles, for some values of $\RL$,
we have also plotted in Figure~\ref{evo} the usual asymptotic diffusion coefficient as given by
Equation~(\ref{Partial}), evaluated after various times. It is clear that it is
consistent with the running diffusion coefficient at large times. However, the running
diffusion coefficient converges at much earlier time, so is therefore more appropriate
for practical evaluation. This is because the running diffusion coefficient depends on
the local (in time) change in the particle mean squared displacement, whereas
Equation~(\ref{Partial}) is a global average and must be evaluated at large times in
order to forget the initial evolution.

To understand the advantages and limitations of the two models of a random magnetic field and
to estimate the numerical resolution required to achieve a desired accuracy,
we compare the diffusion coefficients obtained with the two models.
One might expect that the main deficiency of the magnetic field constructions
will be an insufficient number of modes near the scale
$\kres=2\pi/\RL$, which would likely result in lower than expected levels of
scattering.
Thus, we consider a few values of $\RL$ and a purely random magnetic field, $\eta=1$. Both theory and
earlier numerical results suggest different regimes of diffusion for the cases
$\RL \ll \lc$ and $\RL \gg \lc$, so we consider the cases $\RL=5\lc$, $\lc$ and $\lc/5$, where $\lc$ is obtained from Equation~\eqref{eq:lc}.
Figure~\ref{comparemethods} shows the running isotropic diffusion coefficient
$\kappa\iso=(1/3)\sum_{i=1}^3\kappa_{ii}$ for $s=5/3$. We focus on the time intervals where differences between the
two models are the strongest, so that the relative
differences between the various lines are far larger in Panel (a) than in (b). Note that here
it is necessary to have very good particle statistics so that we can attribute differences
in the diffusion coefficients to changes in the magnetic field resolutions, rather than the statistical
error in the evaluation of a diffusion coefficient; we estimate the maximum relative standard error
in the diffusion coefficients to be less than 1 per cent in all cases.

For $\RL = 5\,\lc$ (Figure~\ref{comparemethods}a), the cosmic rays become diffusive at
$t\approx30\,t\zero$.
In this regime, convergence of the diffusion coefficients in a continuous magnetic field
seems to require a large number of modes $N$ per unit of $\kmax/k\zero$ when compared
with the other panels at lower $\RL$. On the other hand,
large values of $\kmax/k\zero$ may not be required because the particles
are not dominated by magnetic fields at small scales for a sufficiently steep
magnetic spectrum. \citet{PlotnikovEA2011} conclude that $\kmax/k\zero=10$ is adequate for
large $\RL$. For $\kmax/k\zero=96$, the choices $N=150$ and 600 lead to a difference of more
than 10 per cent in the diffusivity. For $N=512$ and $\kmax/k\zero=256$, we
might expect the diffusion coefficient to be smaller than with $N=600$
since the correlation length is smaller (according to the restricted $k$ evaluation of Equation~\ref{eq:lc}). However, the mode density
seems to be too low to realise this with $N=512$, but taking $N=1024$ is sufficient to give the expected result.
For a given $\kmax/k\zero$, we find that the diffusion coefficients always decrease with increasing $N$.
Whereas the continuum model concentrates modes near $k\zero$, the discrete model has very few modes
at low $k$, which is perhaps why we see a large discrepancy between the $512^3$
resolution discrete model and the continuous ones in Panel (a).
By increasing both the mesh separation corresponding to wave number $k\zero$ and
the overall mesh size (denoted by a star
in the figure legend), it appears that
this discrepancy can be eliminated. We tried increasing the resolution alone, up
to $1600^3$, but this had little effect.

In Panel (b), for $\RL = \lc$,
too few modes in the continuous model of Equation~\eqref{meshless} appear to cause a problem.
Overall though, there is less variation in the asymptotic $\kappa$
between various resolutions than in Panel (a).
With the resolution $1024^3$ in the discrete magnetic field model and $N>150$ in the continuous model,
the asymptotic diffusivities agree within about 2 percent over the whole time evolution
(although the $N=600$ case turns out to be somewhat special, leading to a lower value of $\kappa$
than for the other values of $N$). It is also quite satisfactory that significant variation in the resolution
of the discrete model does not affect the result much.

The difference between results obtained under the whole range of magnetic field implementations
is even smaller in Panel (c), for $\RL = \lc/5$. The diffusivities in the three continuous models
coincide almost perfectly, which suggests we have sufficient resonant scattering. It is perhaps surprising that although the discrete model contains many
more modes near $\kres$ than the continuous one, it seems to require that $\kmax$ be several
times larger than $\kres$ for convergence, whereas the continuous one is well converged at $\kmax\approx\kres$. This may be due to the discrete nature of the model
and is discussed further in Section~\ref{section:zeromean}.

The results here suggest that the two models will be in almost perfect agreement in the limit of
high resolution and when $k\zero$ is sufficiently well resolved in the discrete model. Clearly the
discrete model is more limited in the range of $\RL$ that can be considered and presumably the limitations
mentioned above will apply to test particle simulations in more realistic magnetic fields. For the range of $\RL$ studied, we see no
obvious effects due to the periodic magnetic field.
For continuous magnetic field models, some previous works have determined reasonable values of $N$
for particular cases, which has typically been quantified in terms of the number of modes $A$ required
per decade of $\kmax/k\zero$,
i.e. $N = A\log_{10}[\kmax/k\zero]$. Examples of reported reasonable values of $A$ are $25$ \citep{FatuzzoEA2010}, $100$ \citep{Parizot2004}, and $200\text{--}300$ \citep{PlotnikovEA2011}.
Considering the results of figure~\ref{comparemethods}, $A=25$ might be reasonable for $\RL = \lc/5$. However, for $\RL = 5\lc$ this would give $N\approx50$ for $\kmax/k\zero=96$, which would appear to
be unreasonably small. Our results for $N=512, \kmax/k\zero=256$ seem reasonable over all panels, and
this gives $A\approx 213$, which is of similar size to those previously reported. The point to
note here is that a given reasonable value of $A$ is not universally applicable and will depend on both $\RL$ and the desired range $\kmax/k\zero$.

Admittedly we have not performed such a detailed study of the convergence for $\eta < 1$. However, for
 some values of $\eta$, and at the extremes of $\RL$, we have observed similar asymptotic trends to those
 discussed above and in Section~\ref{section:zeromean}, so we suspect that variation of $\eta$ is not
 a very important consideration, except perhaps in the limit $\eta \to 0$.

\label{lastpage}

\end{document}